\documentclass{emulateapj}
\pdfoutput=1
\usepackage{epstopdf}
\usepackage{amsmath}

\usepackage{appendix}
\usepackage{color}

\slugcomment{\today}

\newcommand{\lta}{\lesssim}

\newcommand{\about}{$\sim\!\!$~}
\def\arcdeg{\hbox{$^\circ$}}

\def\arcsec{\hbox{$^{\prime\prime}$}}

\def\snid{\ifmmode{\rm \tt SNID}\else{\tt SNID}\fi}
\def\dm15{\ifmmode{\Delta m_{15}}\else{$\Delta m_{15}$}\fi}
\def\VR{$V\!R$}
\def\magarcsec2{\ \rm{mag
\ arcsec}^{-2}}
\def\ly{\ifmmode{\rm{ly}}\else{ly\fi}}

\shorttitle{Light Echo Profiles \& Spectra}
\shortauthors{Rest et~al.}

\begin{document}

\title{On the Interpretation of Supernova Light Echo Profiles
and Spectra}

\def\harvard{1}

\author{A. Rest\altaffilmark{\harvard,2},
B. Sinnott\altaffilmark{3},
D.~L. Welch\altaffilmark{3},
R.~J. Foley\altaffilmark{4,5},
G. Narayan\altaffilmark{1},
K. Mandel\altaffilmark{4},
M.~E. Huber\altaffilmark{6}
S. Blondin\altaffilmark{7},
}

\altaffiltext{\harvard}{Department of Physics, Harvard University, 17 Oxford Street, Cambridge, MA 02138, USA; arest@stsci.edu}

\altaffiltext{2}{Space Telescope Science Institute, 3700 San Martin
Dr., Baltimore, MD 21218, USA}

\altaffiltext{3}{Department of Physics and Astronomy, McMaster University,
Hamilton, Ontario, L8S 4M1, Canada}

\altaffiltext{4}{Harvard-Smithsonian Center for Astrophysics, 60 Garden Street, 
Cambridge, MA 02138, USA} 

\altaffiltext{5}{Clay Fellow}

\altaffiltext{6}{Department of Physics and Astronomy, Johns Hopkins
  University, Baltimore, 3400 North Charles Street, MD 21218, USA}

\altaffiltext{7}{Centre de Physique des Particules de Marseille (CPPM),
  Aix-Marseille Universit\'e, CNRS/IN2P3, 163 avenue de Luminy, 13288
  Marseille Cedex 9, France}

 
\begin{abstract}
The light echo systems of historical supernovae in the Milky Way and
local group galaxies provide an unprecedented opportunity to reveal
the effects of asymmetry on observables, particularly optical spectra.
Scattering dust at different locations on the light echo ellipsoid
witnesses the supernova from different perspectives and the light
consequently scattered towards Earth preserves the shape of line
profile variations introduced by asymmetries in the supernova
photosphere.  However, the interpretation of supernova light echo
spectra to date has not involved a detailed consideration of the
effects of outburst duration and geometrical scattering modifications
due to finite scattering dust filament dimension, inclination, and
image point-spread function and spectrograph slit width.  In this
paper, we explore the implications of these factors and present a
framework for future resolved supernova light echo spectra
interpretation, and test it against Cas~A and SN~1987A light echo
spectra.  We conclude that the full modeling of the dimensions and
orientation of the scattering dust using the observed light echoes at
two or more epochs is critical for the correct interpretation of light
echo spectra.  Indeed, without doing so one might falsely conclude
that differences exist when none are actually present. 

\end{abstract}

\keywords{supernovae: individual(Cas~A, SN~1987A) ---
supernovae:general --- supernova remnants --- ISM:dust --- methods:
data analysis --- techniques: spectroscopic}

\section{Introduction}
\label{sec:intro}

The first light echoes (LEs) were discovered around Nova Persei 1901
\citep{Ritchey01b, Ritchey01a, Ritchey02, Kapteyn02}. Since then, LEs
(whereby we mean a simple scattering echo rather than fluorescence or
dust re-radiation) have been seen in the Galactic Nova Sagittarii 1936
\citep{Swope40} and the eruptive variable V838 Monocerotis
\citep[e.g.,][]{Bond03}. LEs have also been observed from extragalactic
supernovae (SNe), with SN~1987A \citep{Kunkel87} being the most famous case
\citep{Crotts88, Suntzeff88, Bond90, Xu95, Sugerman05a, Sugerman05b,
Newman06}, but also including SNe~1991T \citep{Schmidt94, Sparks99},
1993J \citep{Sugerman02, Liu03}, 1995E \citep{Quinn06}, 1998bu
\citep{Cappellaro01}, 2002hh \citep{Meikle06, Welch07}, 2003gd
\citep{Sugerman05, VanDyk06}, and 2006X \citep{Wang08}.

The suggestion that historical SNe might be studied by their
scattered LEs was first made by \citet{Zwicky40} and attempted
by \citet{vandenBergh65, vandenBergh66}. The first LEs of
centuries-old SNe were discovered in the Large Magellanic Cloud (LMC)
\citep{Rest05b}. They found three LE complexes associated with
three small and therefore relatively young SN remnants (SNRs) in the
LMC which were subsequently dated as being between 400--1000 years old
using the LMC distance and LE feature positions and apparent motions
\citep{Rest05b}. These findings provided the extraordinary opportunity
to study the spectrum of the SN light that reached Earth hundreds of
years ago and had never been recorded visually or by modern scientific
instrumentation. A spectrum of one of the LEs associated with
SNR~0509-675 revealed that the reflected LE light came from a
high-luminosity SN~Ia \citep{Rest08a}, similar to SN~1991T
\citep{Filippenko92, Phillips92} and SN~1999aa \citep{Garavini04} ---
the first time that an ancient SNe was classified using its LE
spectrum.  Analysis of {\it ASCA} X-ray data of SNR 0509-675 led
\citet{Hughes95} to suggest that this SNR likely came from a SN~Ia.
Recent analysis of its {\it Chandra} X-ray spectra by
\citet{Badenes08} also supported the classification of this object as
a high-luminosity SN~Ia.

The interpretation of LEs, however, is fraught with subtleties.
\citet{Krause05}, for example, identified moving Cas~A features (called
``infrared echoes'' or ``IR echoes'') using mid-IR imaging from the
{\it Spitzer Space Telescope}.  IR echoes are the result of dust
absorbing the outburst light, warming and re-radiating at longer
wavelengths.  The main scientific conclusion of \citet{Krause05} was
that most of the IR echoes were caused by a series of recent X-ray
outbursts from the compact object in the Cas~A SNR based on their
apparent motions.
However, the analysis was flawed because it did not account for the
fact that the apparent motion strongly depends on the inclination of
the scattering dust filament \citep{Dwek08,Rest11_casamotion}.
\citet{Dwek08} also showed that X-ray flares are not energetic
enough to be the source of these IR echoes, but that instead the LEs
must have been generated by an intense and short burst of EUV-UV
radiation associated with the break out of the shock through the
surface of the Cas~A progenitor star. Therefore, the most likely
source of all LEs associated with Cas~A, both infrared and scattered,
is the Cas~A SN explosion itself.  The need to properly analyze the
growing collection of LE spectra is what motivated the present work.

The first scattered LEs of Galactic SNe associated with Tycho's SN and
the Cas~A SN were discovered by \citet{Rest07, Rest08b}.
Contemporaneously, \citet{Krause08a} obtained a spectrum of a
scattered optical LE spatially coincident with one of the Cas~A IR
echoes, and identified the Cas~A SN to be of Type~IIb.  Because
\citet{Rest08b} discovered several LEs with multiple position angles
relative to the SNRs, there is a new opportunity for observational SN
research --- the ability to measure the spectrum of the same SN from
several different directions. At any given instant after the SN light
has reached an observer by the direct path, an ellipsoidal surface
exists where scattered light from the SN may reach the observer, whose
arrival to the observer is delayed by the additional path
length. Interstellar dust concentrations must lie at one or more
locations on the ellipsoid for the scattered LE to be detectable.
Since each LE is at a unique position on the ellipsoid, each LE has a
unique line of sight to the SN.  Therefore, the scattered spectral
light can provide spectral information on the asymmetry of the SN
photosphere. In the limiting case of dust on the ellipsoid opposite
the observer, the spectrum of the scattered LE carries the signature
of conditions on the SN hemisphere usually hidden from the
observer. As the time since the explosion increases, the scattering
ellipsoid surface expands and may intersect additional dust
concentrations. We present a framework on how to interpret observed LE
spectra depending on the scattering dust properties, seeing, and
spectrograph slit width and position.

In this paper, we examine the pitfalls of previous analyses, suggest
improvements for future studies, and present the application of our
methods to data.  In Section~\ref{sec:case}, we describe how the
properties of the scattering dust can cause significant differences to
observed LE spectra, and show a specific example of how this has been
overlooked in the past.  In Section~\ref{sec:analysis}, we quantify
how observed LE spectra depend on potential differences in dust
characteristics (thickness and inclination), as well as observational
characteristics (seeing and spectroscopic slit size and orientation).
We then apply this framework to observations of Cas~A
(Section~\ref{sec:casa}) and SN~1987A (Section~\ref{sec:87A}).  In
Section~\ref{sec:discussion}, we discuss how differences in the dust
and observing conditions can effect an analysis of LE spectra and
suggest future applications of LE observations.  We conclude in
Section~\ref{sec:conclusions}.

\section{A Case Study: The Krause et~al.\ (2008) Cas~A Light Echo
Spectrum}
\label{sec:case}

\begin{figure*}
   \epsscale{1.15}
   \plotone{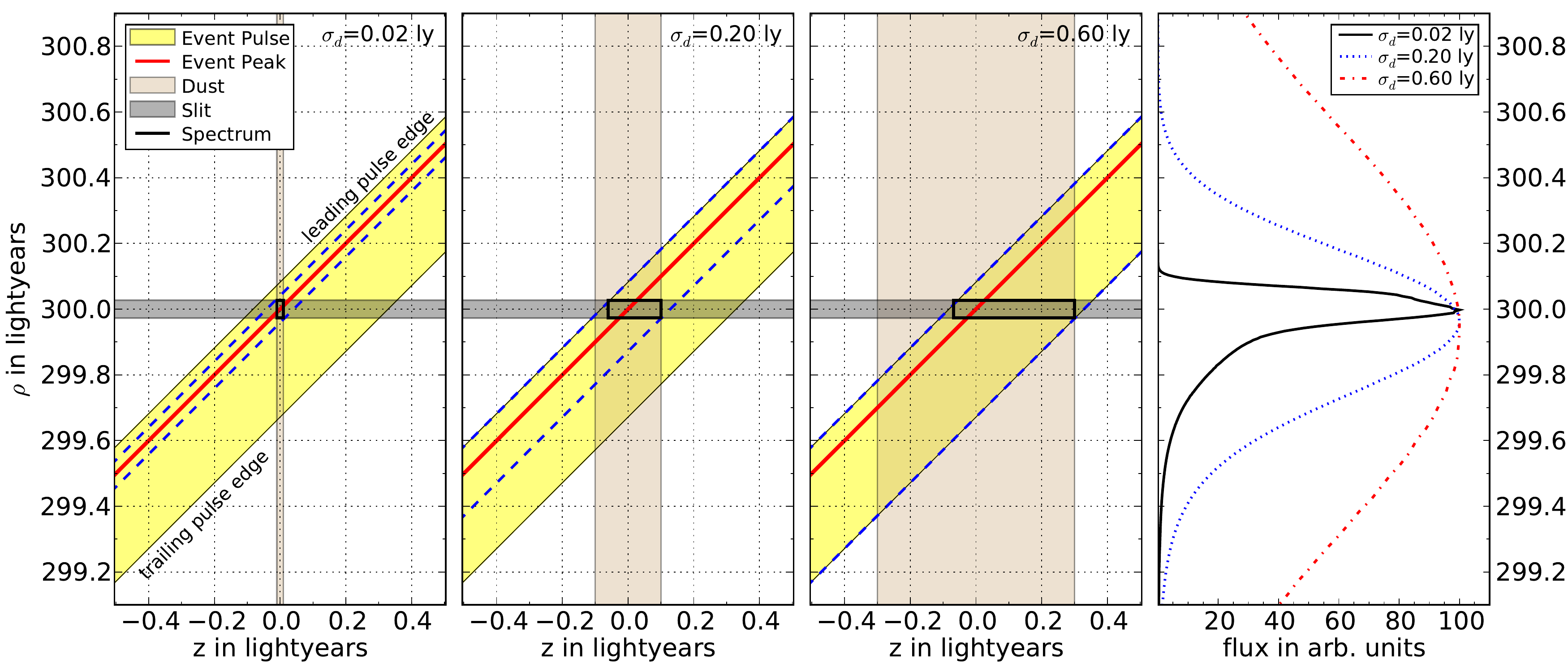}

   \caption[]{Illustration of how the dust filament width,
     $\sigma_{d}$, influences the observed LE spectrum.  The yellow
     shaded area indicates a 300-year-old SN event that lasts 150
     days, with a rising and declining arm of 30 and 120 days,
     respectively. The peak is indicated with a red line.  The
     scattering dust filament (light grey shaded region) is located in
     the plane of the sky crossing the event position ($z = 0.0$~ly),
     with an inclination of $\alpha = 0$\arcdeg.  The dust width is
     0.02~ly, 0.2~ly, and 0.6~ly from left to right, respectively, and
     the left panel shows the respective LE profiles.  A spectroscopic
     slit with a width of $1\arcsec$ is indicated with the dark gray
     shaded area. Only light that is within the intersection of the
     event pulse, dust, and slit (indicated with the thick black
     rectangle) contributes to the observed spectrum. The blue dashed
     lines indicate which part of the light pulse is probed by the
     spectrum.
\label{fig:illustration_dustwidth}}
\end{figure*}

A common assumption is that an observed LE spectrum is equivalent to
the light-curve weighted integration of the spectra at individual
epochs \citep[e.g.,][]{Rest08a,Krause08a}.  However,
Figure~\ref{fig:illustration_dustwidth} illustrates how slit width,
dust filament width and inclination significantly influence the
observed LE spectrum. We consider a LE from a 300-year-old event
10,000 light years away that lasts 150 days (yellow shaded area), with
a rising and declining arm of 30 and 120 days, respectively. The peak
of the event is indicated with the red line. We place the origin of
our coordinate system $\rho,z$ at the source event. The $z$-axis is
the distance of the dust to the source event along the line of sight.
The $\rho$-axis is in the plane of the sky from the source event to
the LE position. The quadratic relation between the parameters $z$,
$\rho$, and the time since explosion $t$ is described by the
well-known LE equation \citep{Couderc39} (see
Section~\ref{app:leprofile} for details).

The scattering dust filament (brown shaded area) is located at $(\rho,
z) = (300, 0)$~ly with an inclination of $\alpha = 0$\arcdeg\ where
$\alpha$ is the angle with respect to the plane of the sky (see
Equation~\ref{eq:dustsheet}). The dust width is 0.02~ly, 0.2~ly, and
0.6~ly from left to right, respectively.  The rightmost panel of
Figure~\ref{fig:illustration_dustwidth} shows the flux of the LE that
we would observe with these dust filaments versus $\rho$. We denote
these as the LE profiles, which essentially are cuts through the LE
along the axis pointing toward the source event. In order to be able
to directly compare the LE profiles to the dust and LE pulse geometry,
we show the flux and $\rho$ as the x-~and~y-axis, respectively. Note
that for very thin dust filaments like the one with $\sigma_d =
0.02$~ly, the LE profile is the the projected light curve.

A spectroscopic slit with a width of $1\arcsec$ is indicated with the
gray shaded area. In the leftmost panel, the dust filament has a dust
width of 0.02~ly (0.006~pc).  Only light that is in the intersection
of the event pulse, dust, and slit (indicated with the thick black
rectangle) contributes to the observed spectra. In this case, only
20\% of the event (30 out of 150 days, indicated by the dashed blue lines)
contribute to the spectrum. The second panel from the left shows the
same example, but this time the dust filament has a width of 0.2~ly
(0.06~pc), and \about 51\% of the event pulse (77 out of 150 days)
contributes to the spectrum. Only if the dust width is at least 0.6~ly
(0.18~pc) thick (third panel from the left), the full event
contributes to the spectrum.

\begin{figure}
\epsscale{2.3}
\plottwo{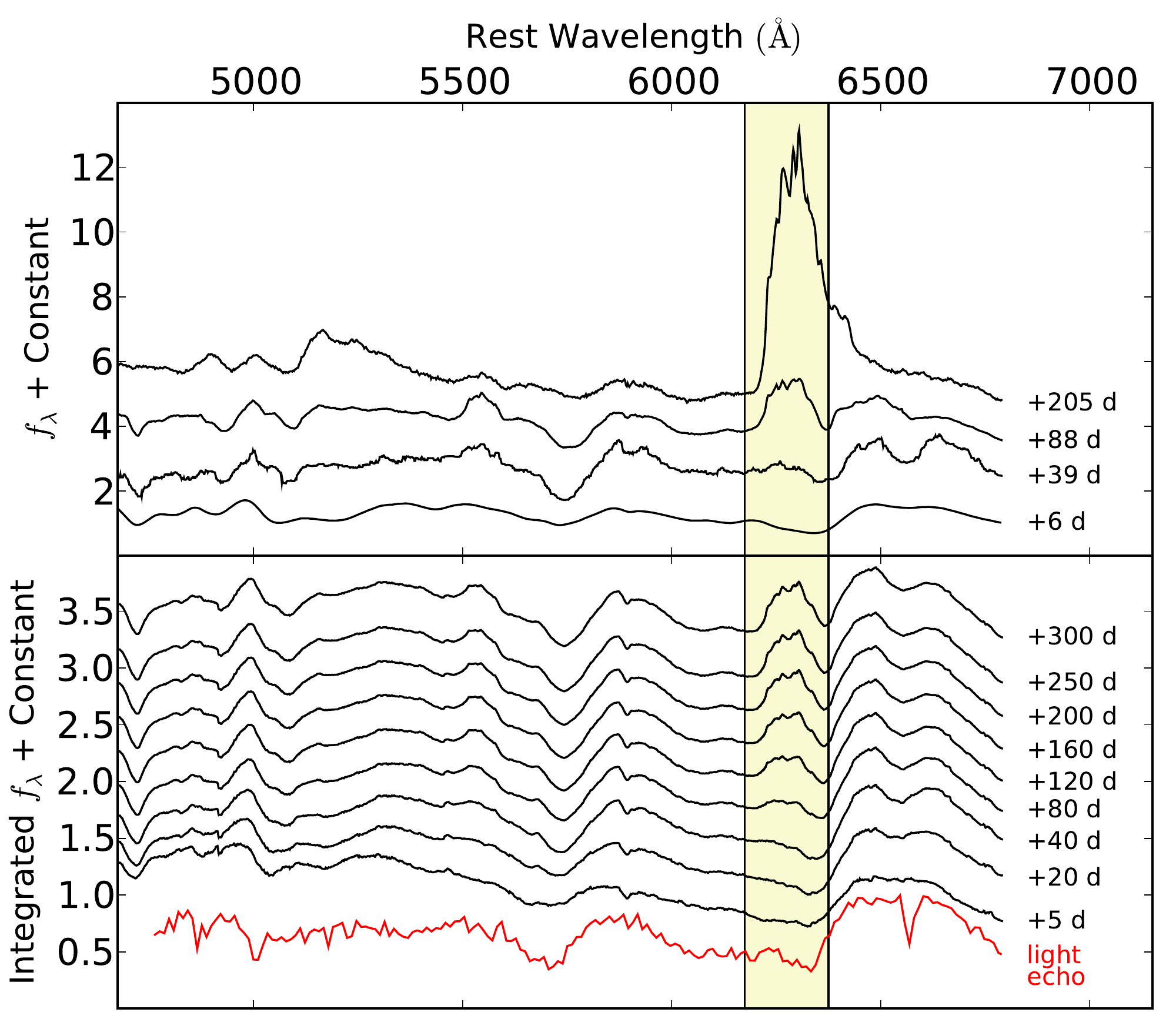}{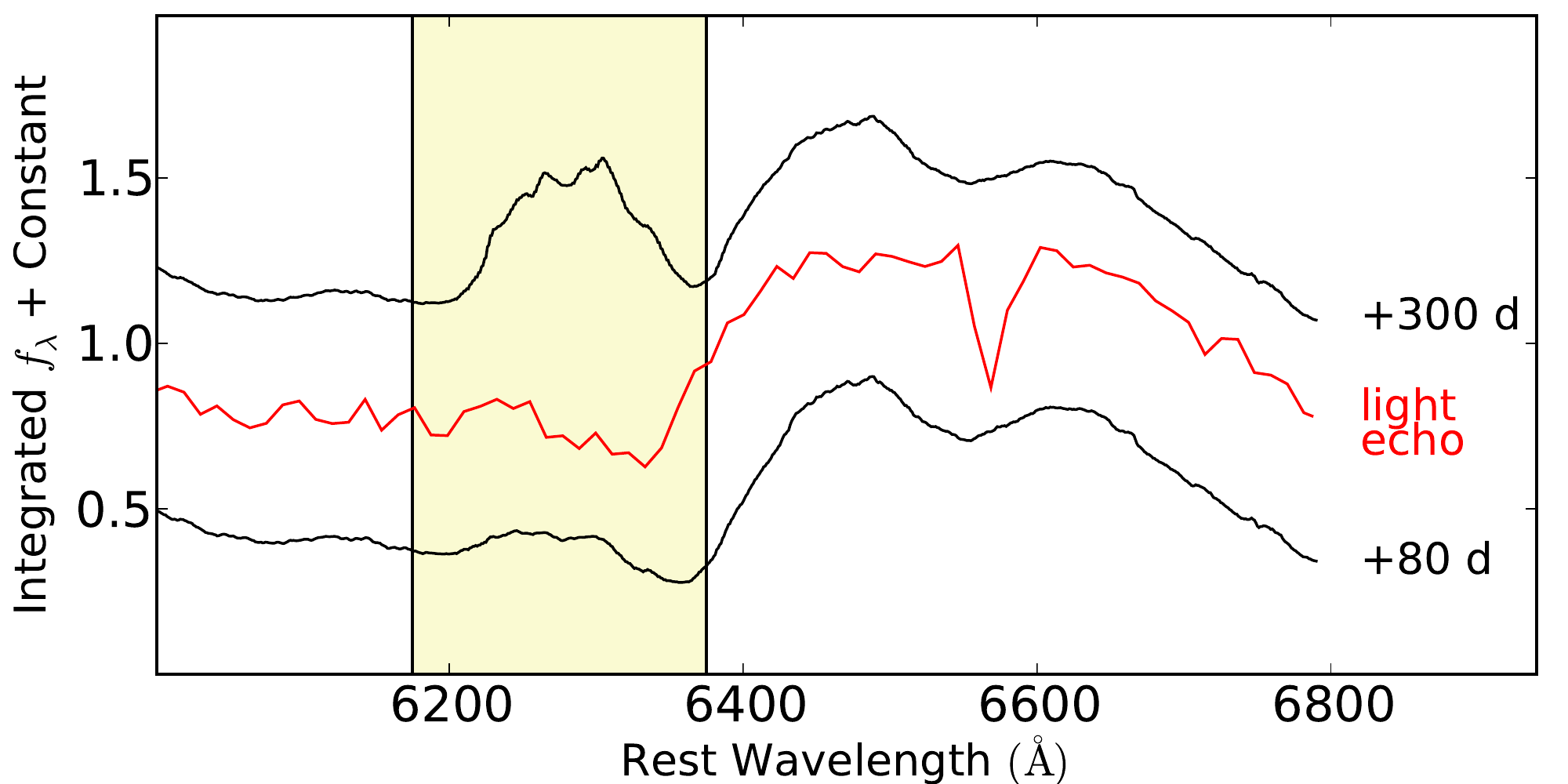}
\caption[]{\emph{Top panel:} Spectra of SN~1993J at $+6$, $+39$,
$+88$, and $+205$~days relative to peak brightness from the bottom to
the top, respectively.  During the late phases ($\gtrsim 40$~days
after peak), a significant fraction of flux is emitted in the
[\ion{O}{1}] $\lambda\lambda 6300$, 6364 doublet, which dominates the
spectra after \about 200~days.  The [\ion{O}{1}] doublet is indicated
by the yellow-shaded region.
\emph{Middle panel:} light-curve weighted integrated spectra of
SN~1993J. The integration limits of these spectra are from $-20$~days
relative to peak brightness to $P_{\max}$, where $P_{\max}$ ranges
from $+5$ to $+300$~days from the bottom to the top.  The most
significant difference between the integrated spectra with early and
late integration limits is the emergence of the [\ion{O}{1}]
doublet. The Cas~A LE spectrum of \citet{Krause08a} is shown in red,
which does not show a prominent [\ion{O}{1}] doublet feature.
\emph{Bottom panel:} light-curve weighted integrated spectra of
SN~1993J in the wavelength region near the [\ion{O}{1}] doublet and
H$\alpha$ features.  The integrated spectra have integration limits of
$+80$~days and $+300$~days are compared to the Cas~A LE spectrum of
\citet[red curve]{Krause08a}.  The integrated SN~1993J spectrum with
the $+80$~day integration limit is a better fit to the [\ion{O}{1}]
doublet in the observed Cas~A spectrum.
\label{fig:sn93j}}
\end{figure}

This clearly demonstrates that the observed LE spectrum is not
necessarily the full light-curve weighted integrated spectrum when the
scattering dust filament is sufficiently thin. The effect of a thin
filament of dust on the interpretation of a LE spectrum is best
appreciated by considering the difference in persistence of spectral
features at early and late time. Naively, one might expect that the
faintness of a SN at late times would not result in spectral features
from those phases contributing to the integrated spectrum in a
significant way.  However, in the late nebular phases, most of the
flux is typically concentrated in very few lines. The top panel of
Figure~\ref{fig:sn93j} portrays the spectra of SN~1993J, the very
well-observed and prototypical example of the SN~IIb class
\citep[e.g.,][]{Filippenko93, Richmond94, Filippenko94, Matheson00},
at 6, 39, 88, and 205~days after peak brightness from the bottom to
the top, respectively \citep{Jeffery94,Barbon95,Fransson05}.  During
the late phases, a significant fraction of flux is radiated in the
[\ion{O}{1}] $\lambda\lambda 6300$, 6364 doublet, which completely
dominates the spectra for times later than 200 days.  The bottom panel
shows the light-curve weighted integrated spectra of SN~1993J with
different integration limits. The integration limits of these spectra
are from 20 days before peak brightness to $P_{\rm max}$, where
$P_{\rm max}$ ranges from 5 to 300~days after peak brightness from the
bottom to the top. The most significant difference between the
integrated spectra with early and late integration limits is the
emergence of the [\ion{O}{1}] doublet. We use the SN~1993J lightcurve
from \citet{Richmond96} as weights for the integration.

SNe~IIb are not unique in this regard.  The nebular spectra of all SNe
are dominated by strong emission lines.  Specifically, SNe~Ia have
strong [\ion{Fe}{2}] and [\ion{Fe}{3}] lines, SNe~Ib/c have strong
[\ion{O}{1}], [\ion{Mg}{1}], \ion{Ca}{2}, and [\ion{Ca}{2}] lines, and
SNe~II have strong hydrogen Balmer lines.  See \citet{Filippenko97}
for a review of SN spectroscopy.

The perils of not properly accounting for differences in dust filament
thickness are clear --- the strength of features in observed LE
spectra can be significantly altered by the width
of the scattering dust on the sky.

In Figure~\ref{fig:sn93j}, we display the LE spectrum of Cas~A by
\citet{Krause08a} in red.  This spectrum has a weak [\ion{O}{1}]
doublet, consistent with the integrated spectra of SN~1993J, but {\it only
if integrated until \about 80~days after peak brightness.}  The full
integrated spectra of SN~1993J have a much stronger [\ion{O}{1}]
feature.  If the LE spectrum is the full light-curve weighted
integrated spectrum of Cas~A, then Cas~A and SN~1993J have a
significant difference at wavelengths near the [\ion{O}{1}] feature.
The alternative is that the LE spectrum does not contain light from
the late-time portion of the SN light curve, i.e., the weighting
function is truncated at late times.

In their comparison to the observed Cas~A LE, \citet{Krause08a}
also used an integrated SN~1993J spectrum --- but only integrated to
83~days after peak brightness.  This seemingly prudent choice was not
explained in the text, but presumably they found this integration time
to yield the best comparison.  However, this analysis leaves an open
question:  Was the Cas~A SN different from SN~1993J in this regard or
are the differences due to the properties of the scattering dust?  In
the following sections, we attempt to answer this question and its
more general counterpart for all LEs.

\section{Light Echo Profile Modeling}
\label{sec:analysis}

In Appendix~\ref{app:leprofile}, we derive that the observed LE
profile shape, $P$, strongly depends on the dust thickness, dust
inclination, and seeing. We also show that this may have a significant
impact on the observed LE spectra (see
Appendix~\ref{app:lespec}~and~\ref{app:slitmisalignement}). Here we
investigate the magnitude of each of these effects on observed LE
spectra.

For this analysis, we examine a toy model of a LE that, despite its
simplicity in some respects, appears to be representative of actual LE
situations. The toy model is a LE from a 300-year-old event 10,000~ly
from the observer. For the event light curve, we choose a ``sail''
shape, with the peak at phase 0.0 years and then decreasing to zero
flux at a phase of 0.3~years and $-0.1$~years (see the dotted curves
in the right-hand panel of Figures~\ref{fig:thickness_changes} through
\ref{fig:slitoffset_changes_seeing1arcsec} for a representation of
this light curve), 
sharing the characteristic short rise and long decline
of a SN light curve. 
The scattering dust filament intercepts the LE paraboloid at
$(\rho_{0},z_{0}) = (300,0)$~ly.  For the dust model, we use a dust
filament that is locally planar and perpendicular to the $(\rho,z)$
plane, with a boxcar width profile.  In the following subsections, we
vary the properties of a single parameter while holding the others
constant.  We are therefore able to determine the effects of each
parameter on the observed LE and its spectrum.

\subsection{Dust Thickness}
\label{sec:thickness_changes}

As described above, the dust thickness can significantly alter the LE
spectrum.  Using our toy model, we can quantify this effect.

\begin{figure}
\epsscale{1.2}
\plotone{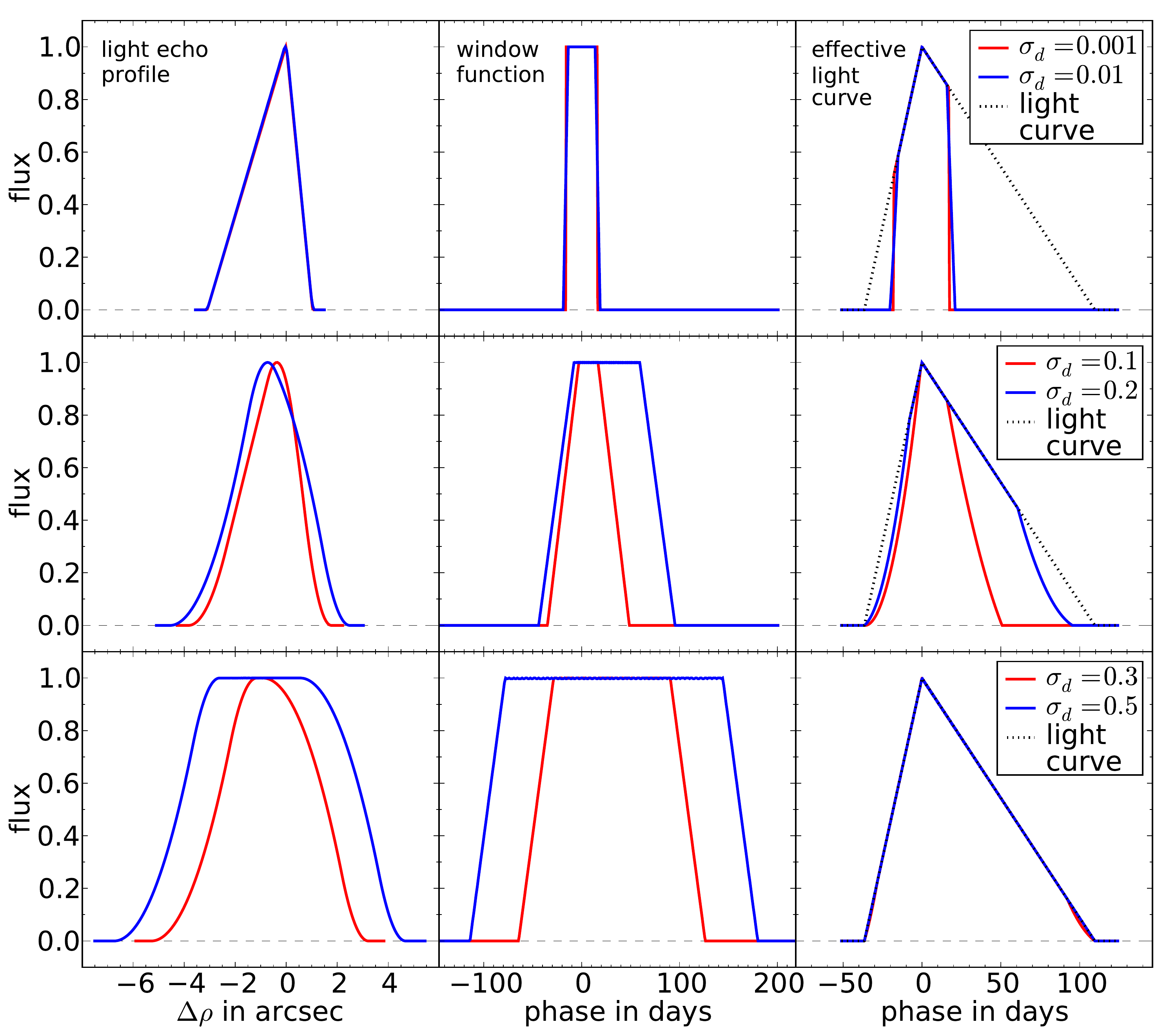}
\caption[]{Modeled LE profiles (left column), window functions (middle
column), and effective light curves (right column) for our toy model
as a function of dust width.  All models in all rows have the same
parameters except for the dust width; the separation into three rows
is only for clarity.  $\Delta \rho$ is with respect to the position
where the dust intersects the light pulse.  The toy model is a
300-year-old event with a ``sail'' shape as described in the text.
The scattering dust filament intersects the event pulse paraboloid at
$z = 0.0$, and has an inclination of $\alpha = 45$\arcdeg.
\label{fig:thickness_changes}}
\end{figure}

Figure~\ref{fig:thickness_changes} presents LE profiles, window
functions, and effective light curves for the toy model with different
dust widths.  The left column shows the LE profiles for each dust
thickness.  The $x$-axis is $\Delta \rho = \rho - \rho_{0}$, where
($\rho_{0}, z_{0}$) is the position where the dust intersects with the LE
paraboloid at phase 0.0 years of the event.  We choose a dust filament
inclination of $\alpha = 45$\arcdeg, since it is perpendicular to the LE
paraboloid and thus most representative of typical inclinations.
Centering a $1.0\arcsec$ spectroscopic slit on the peak of the LE
profile, we calculate the associated window function (middle column)
and effective light curves (right column), which are simply the light
curve convolved with the window function.  The different rows show
different dust widths.

There are two distinct regimes where a modest change in the dust width
has little effect on the effective light curve or integrated spectrum.
The first one is shown in the top panels, where the dust width is so
thin that the observed LE profile is effectively the projected initial
light curve (corresponding to the left panel of
Figure~\ref{fig:illustration_dustwidth}).  The other limit is a very
thick dust sheet (corresponding to the right panel of
Figure~\ref{fig:illustration_dustwidth}).  


In the former case (top row of Figure~\ref{fig:thickness_changes}),
changing the dust width from $\sigma_{d} = 0.001$~ly to $\sigma_{d} =
0.01$~ly only marginally modifies the effective light curve. Centering
a spectroscopic slit on the LE profile results in a box-car window
function, and the effective light curve is truncated to the region
near maximum brightness.  In the latter case (bottom row of
Figure~\ref{fig:thickness_changes}), the dust width is large enough
that the LE profile is dominated by the dust width.  The LE profile
has a flat top, and only the edges are impacted by the original light
curve profile.  Centering a spectroscopic slit on the LE profile
results in very wide window functions, and the effective light curve
is essentially the same as original light curve (see the right panel
of Figure~\ref{fig:illustration_dustwidth} for an illustration).

As we show in Sections~\ref{sec:casa} and \ref{sec:87A}, our
observations of LEs have been a mixture of both regimes. The middle
row of Figure~\ref{fig:thickness_changes} shows the light curves,
window functions, and effective light curves for dust widths of
$\sigma_{d} = 0.1$~ly (0.03~pc) and $\sigma_{d} = 0.2$~ly (0.06~pc).  The
effective light curve is still heavily weighted to the time near
maximum\footnote{In this example, with increasing dust width, the peak
of the LE profile moves away from $\Delta \rho$ towards the negative
direction.  This is caused by using a box-car profile for the dust
width profile. This effect is minimized if the dust width profile is
peaked (e.g., if it is a Gaussian).}, but with increasing dust width
the effective light curve slowly transitions to the full light curve.

\subsection{Dust Inclination}
\label{sec:alpha_changes}

The dust inclination has a large effect on the LE profile.  This is
simply a projection effect; as the dust sheet rotates, the projection
of the light curve on the sky can change significantly.  Turning to
our toy model, we quantify this effect.

\begin{figure}
\epsscale{1.2}
\plotone{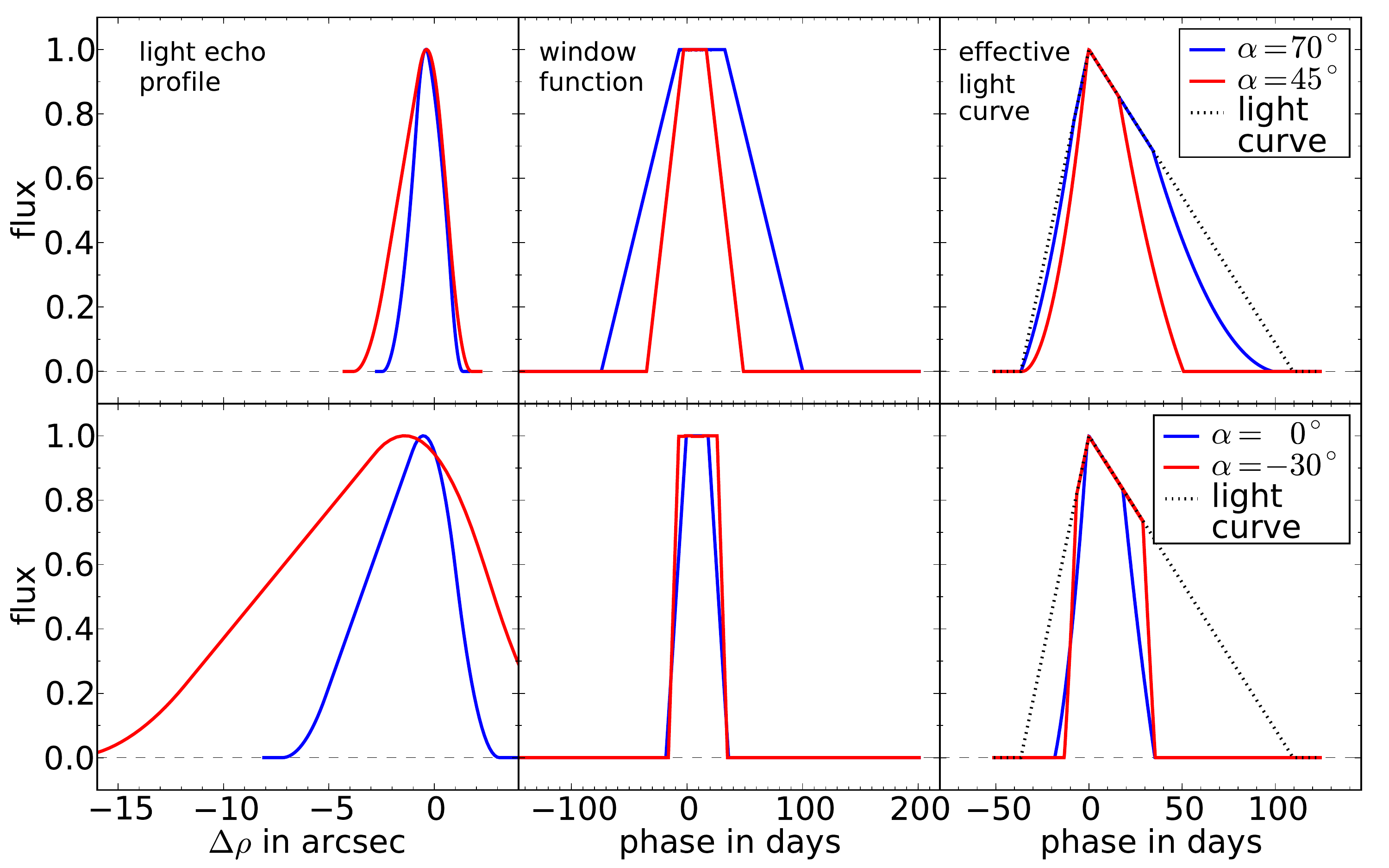}
\caption[]{Same as Figure~\ref{fig:thickness_changes}, but
for varying dust filament inclinations, $\alpha$, and constant dust
width.  The width of the dust filament is $\sigma_{d} = 0.1$~ly.  Note
that the LE profile gets ``stretched'' when the inclination of the
dust filament approaches the tangential angle of the LE ellipsoid,
which is $\alpha = -45$\arcdeg\ for $z = 0$~ly.
\label{fig:alpha_changes}}
\end{figure}

We fix the dust width to $\sigma_{d} = 0.1$~ly.
Figure~\ref{fig:alpha_changes} shows modeled LE profiles, window
functions, and effective light curves for several examples that only
differ by the inclination of the scattering dust filament.  As
expected, we find that the filament inclination has a profound impact
on the LE profile width.  For inclinations close to $90$\arcdeg\ (see
$\alpha = 70$\arcdeg\ in the top row of
Figure~\ref{fig:alpha_changes}), the dust filament is aligned along
the line of sight, while the light curve projected into $\rho$ space
is compressed. Consequently, the window function has large wings, and
the effective light curve is very similar to the full light curve.
However, for inclination angles corresponding to dust sheets close to
the tangent plane of the LE paraboloid (for $z_{0} = 0.0$~ly is
$-45$\arcdeg), the LE profile subtends a larger angle on the sky (see
$\alpha = -30$\arcdeg\ in bottom row of
Figure~\ref{fig:alpha_changes}).  As a result, a spectroscopic slit
will cover a very small portion of the light curve, and the window
function is approximately a box car with a small width.

It is clear that the determining the dust inclination is very
important, but fortunately it is straightforward to do so from the
observations in cases where the location of the SN and date of
outburst are known. In Appendix~\ref{app:inclination} and in more
detail in Rest et~al.\ (in prep.), we derive the dust inclination from
the apparent motion of the LE in difference images, and discuss its
sources of uncertainty.
%
%
%
%
We discuss the impact the dust inclination has on the possibility of
sampling the spectra within different age intervals for the same SN, as
well as the possibility of constraining the SN light curve shape in
Section~\ref{sec:casa_timeresolve}~and~\ref{sec:casa_constraininglc},
respectively.

\subsection{Point-Spread Function Size}
\label{sec:psfsize_changes}

Observations of resolved sources must always account for the
point-spread function (PSF).  LEs are no exception since degradation
of an image by a large PSF will not only smear the LE across the sky,
but will also smear the LE in {\it time}.  Again, we turn to our toy
model, and vary the PSF to determine its effect on the interpretation
of LEs.

\begin{figure}
\epsscale{1.2}
\plotone{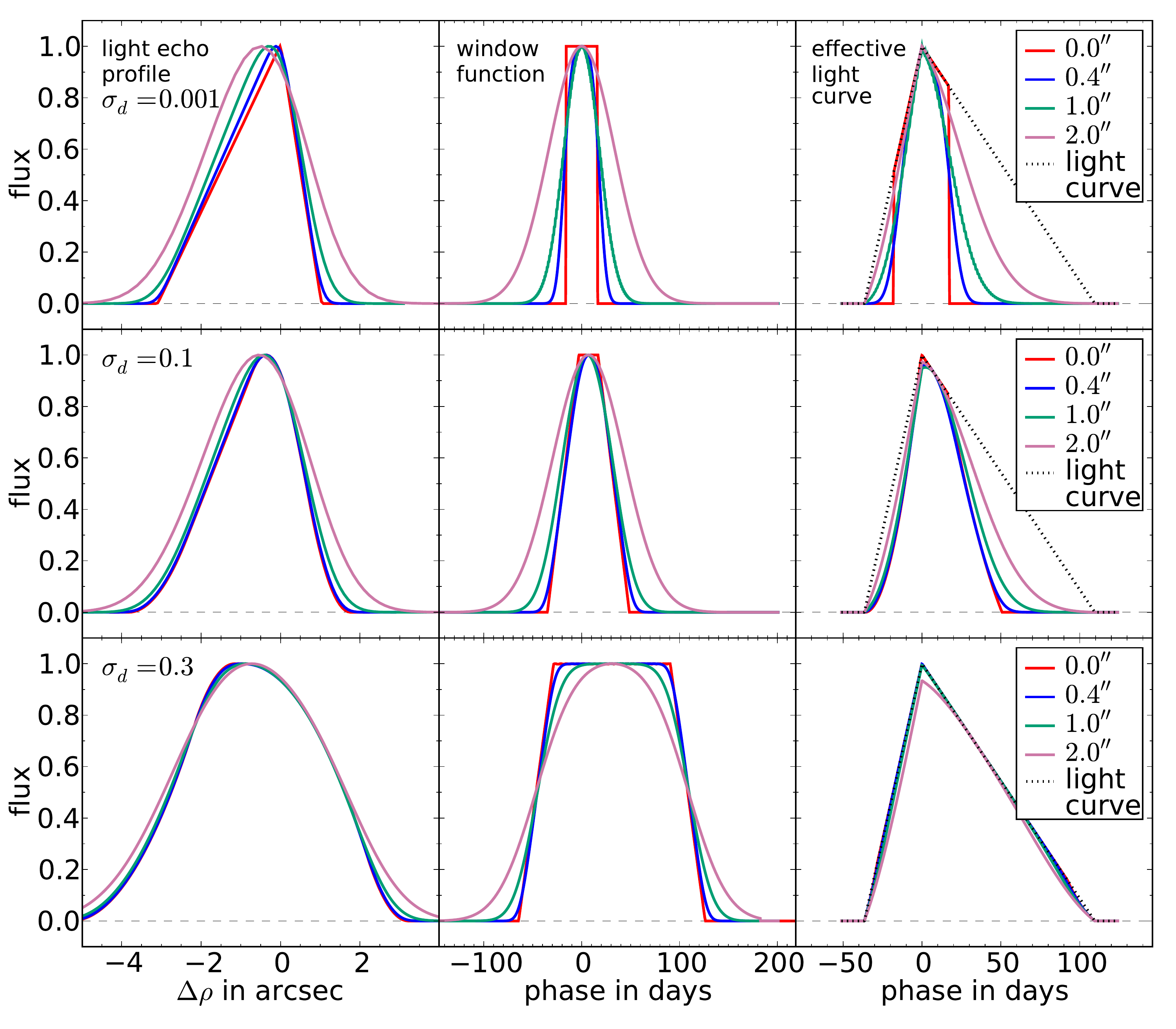}
\caption[]{Same model as in Figure~\ref{fig:thickness_changes}, but
with varying the dust filament width and seeing.  The dust width
presented are $\sigma_{d} = 0.001$~ly (top row), $\sigma_{d} = 0.1$~ly
(middle row), and $\sigma_{d} = 0.3$~ly (bottom row).  Seeing of
$0.0\arcsec$, $0.4\arcsec$, $1.0\arcsec$, and $2.0\arcsec$ is
indicated with red, blue, yellow, and green lines, respectively.
\label{fig:seeing_changes}}
\end{figure}

Figure~\ref{fig:seeing_changes} shows our toy model with dust
inclination of $\alpha=45$\arcdeg.  We examine a Gaussian PSF with four
different full-widths at half-maximum (FWHMs): $0.0\arcsec$
(corresponding to a $\delta$ function PSF), $0.4\arcsec$, $1\arcsec$,
and $2\arcsec$.  Each row of Figure~\ref{fig:seeing_changes}
represents a different dust thickness, with the top, middle, and
bottom rows corresponding to $\sigma_{d}=0.001$, $0.1$, and $0.3$~ly,
respectively.  For small dust widths ($\lta 0.1$~ly), and thus thin LE
profiles, the effective light curve depends significantly on the
seeing.  However, for larger dust widths, the seeing has very little
effect on the effective light curve.  This insight provides guidance
for prioritizing targets for spectroscopic follow-up observations based on
the conditions at the telescope.

\subsection{Slit Offset}
\label{sec:slitoffset}

\begin{figure}
\epsscale{1.2}
\plotone{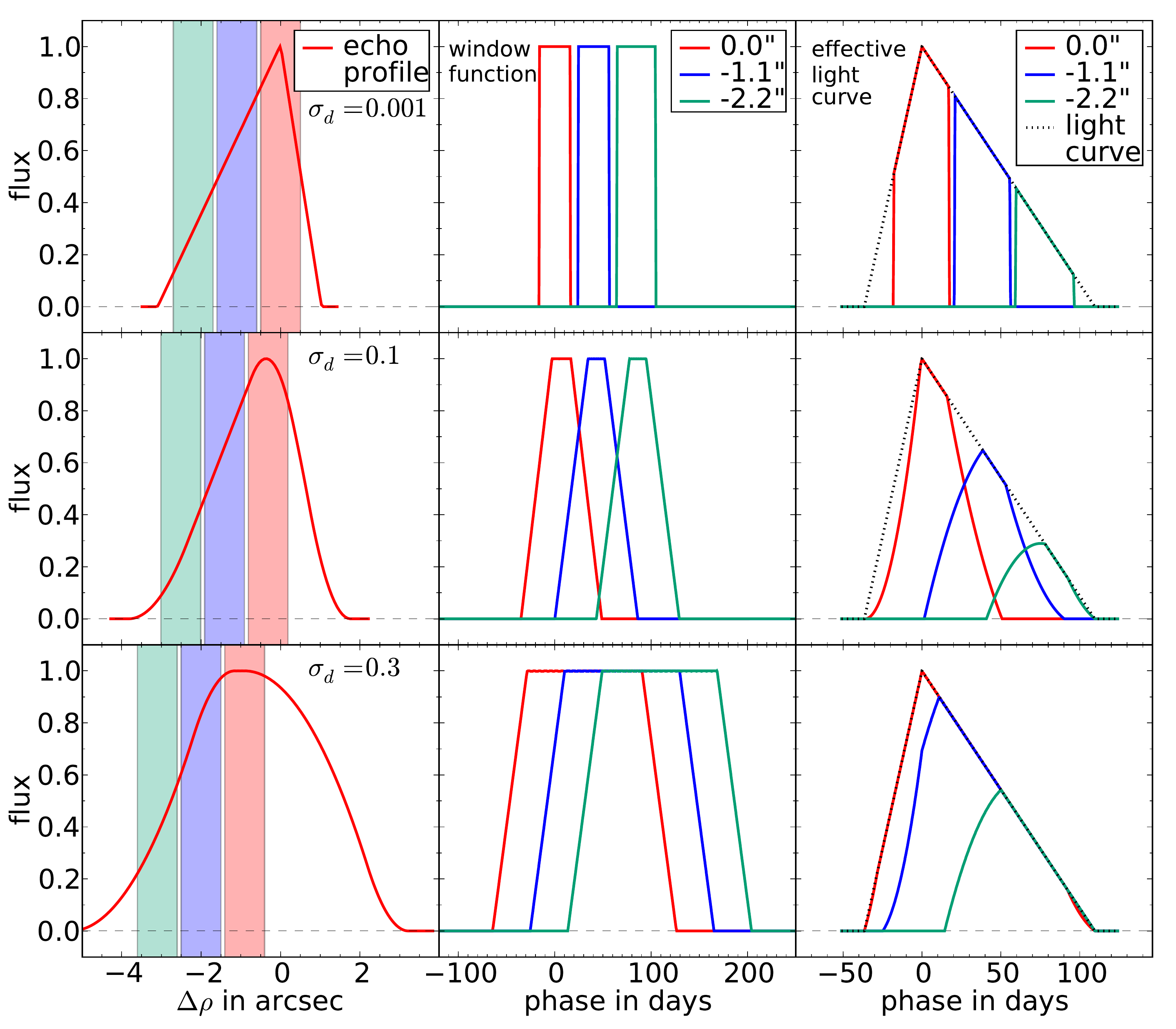}
\caption[]{Same as Figure~\ref{fig:thickness_changes}, but with slits
that are offset by $0.0\arcsec$, $-1.1\arcsec$, and $-2.2\arcsec$ from
the peak of the LE profile. The $1\arcsec$ slits are indicated with
the shaded areas in the left column.
\label{fig:slitoffset_changes}}
\end{figure}

\begin{figure}
\epsscale{1.2}
\plotone{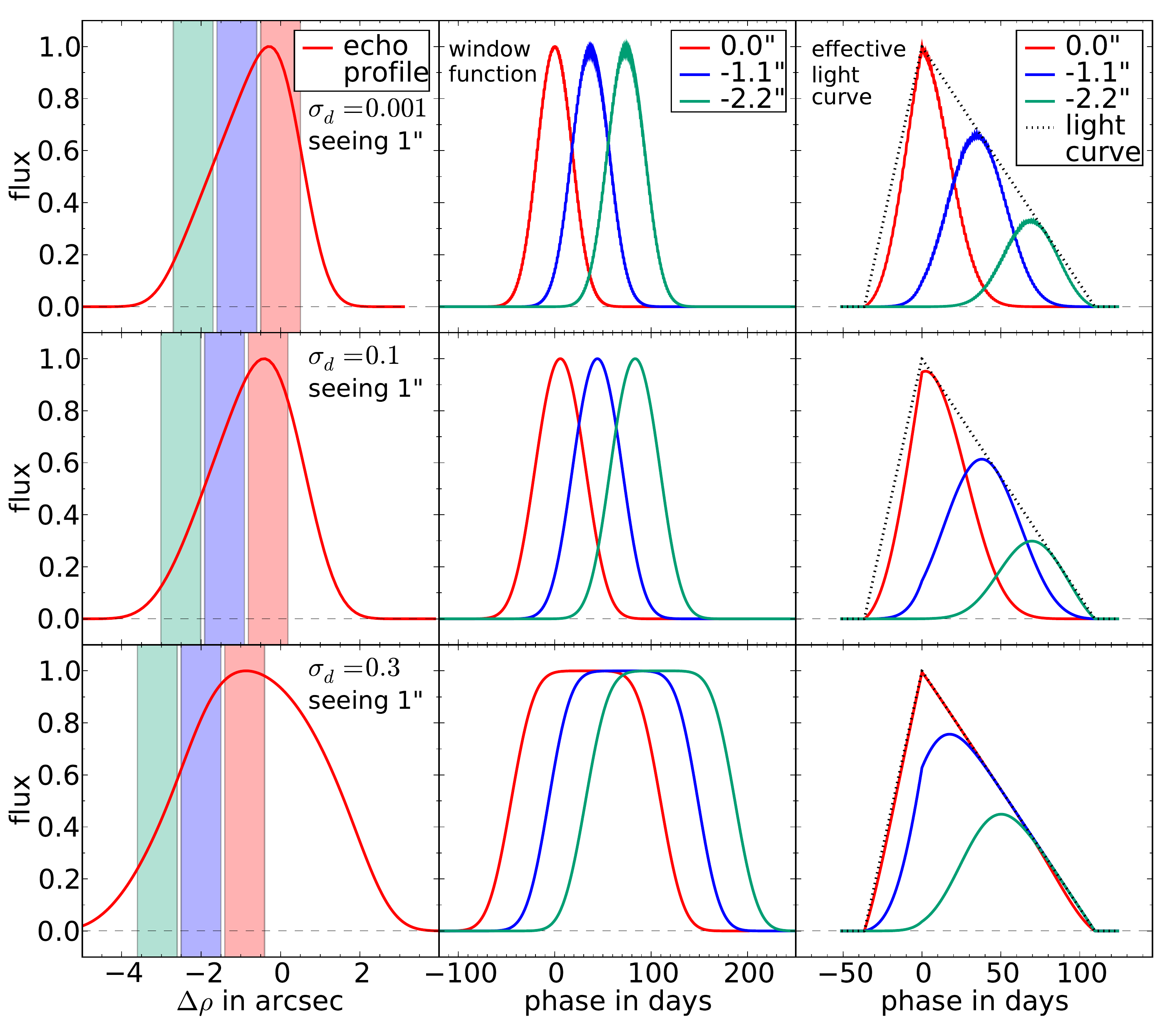}
\caption[]{Same as Figure~\ref{fig:slitoffset_changes} but in
$1\arcsec$ seeing.
\label{fig:slitoffset_changes_seeing1arcsec}}
\end{figure}

Thus far, we have assumed for our toy model that the spectroscopic
slit was centered on the peak of the LE profile.  However, LEs are
usually both very faint and extended.  Additionally, their apparent
motion on the sky makes perfect alignment of a spectroscopic slit
unlikely - especially when there is a significant delay between the
time of mask preparation and observations.  
Figure~\ref{fig:slitoffset_changes} shows the effect of
having the slit offset from the LE profile peak.  As before, we choose
a representative dust inclination of $\alpha = 45$\arcdeg, and the
three rows, from top-to-bottom, correspond to dust widths of $\sigma_{d} =
0.001$, $0.1$, and $0.3$~ly, respectively.  The red, blue, and green
shaded areas indicate $1\arcsec$ slits with $0.0\arcsec$,
$-1.1\arcsec$, and $-2.2\arcsec$ offsets from the peak, respectively.
Clearly a $1\arcsec$ offset has a significant impact on the effective
light curve for all but the thickest dust widths (but even in that
case, it could still result in a significant difference if the 
spectroscopic features in the outburst spectra are changing 
rapidly with time).
Figure~\ref{fig:slitoffset_changes_seeing1arcsec} shows the same
scenario as Figure~\ref{fig:slitoffset_changes}, but with the addition
of a $1\arcsec$ PSF to the model.

\begin{figure}
\epsscale{1.2}
\plotone{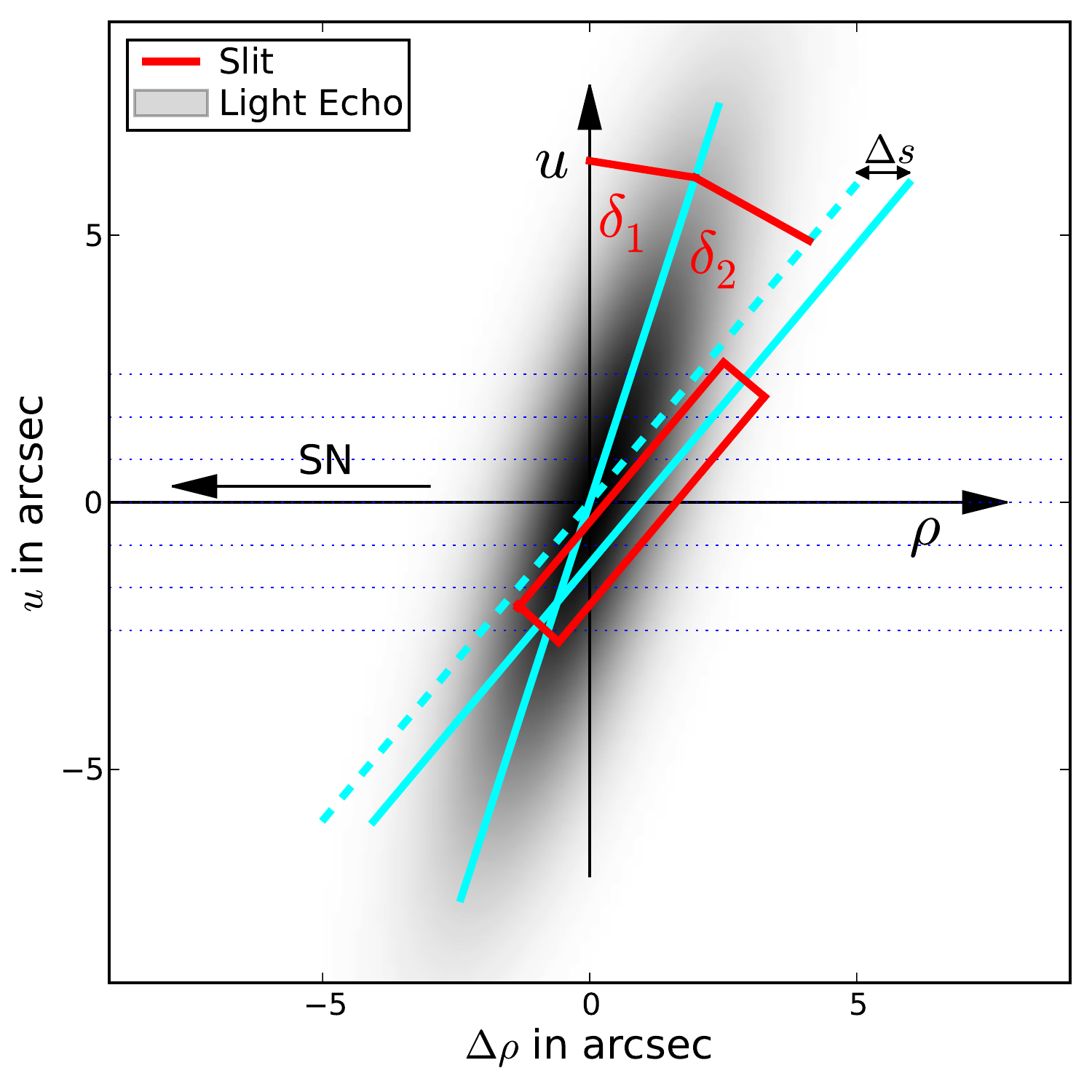}
\caption[]{Toy model of a LE and spectroscopic slit configuration
where the slit is misaligned both in angle and offset relative to the
LE.  The LE is indicated with the shaded area.  We define a coordinate
system ($\rho, u$) in the plane of the sky centered at the LE, where
the $\rho$ axis is in the direction to the source event, and $u$ is
perpendicular to $\rho$, with the positive direction along positive
position angle of LE-SNR. The LE is along the $u$ axis if the dust
filament is perpendicular to the ($\rho, z$) plane. In this example he
angle $\delta_1$ between the LE and the $u$-axis is 18\arcdeg, and the
angle $\delta_2$ between slit and LE is 22\arcdeg.  Six subslits of
width $0.8\arcsec$ are indicated with the blue dotted lines.  The
offset $\Delta s$ is $1.0\arcsec$.
\label{fig:slitpos_illu}}
\end{figure}

Figure~\ref{fig:slitpos_illu} illustrates the application of this
model to a ``real world'' set of conditions.  We define a coordinate
system ($\Delta\rho,u$), where as before $\Delta\rho = \rho-\rho_{0}$,
and $u$ is orthogonal to the $\rho$-axis through $\Delta\rho=0.0$ in
the plane of the sky. For a dust filament that is locally planar and
perpendicular to the ($\rho,z$) plane, i.e. parallel to $u$, the LE is
aligned with the $u$ axis. However, if it is tilted with respect to
$u$ by an angle of $\delta_1$, then the LE will also be tilted by the
same angle (see gray shaded area in illustration). We consider a slit
(red box in Figure~\ref{fig:slitpos_illu}) that is misaligned with the
LE by an angle $\delta_2$, and also offset along the $\rho$ axis by
$\Delta s$.

In this situation, we can recover the proper LE window functions.  To
do so, we can split the slit into multiple subslits, (in this example,
we show this with the horizontal dotted lines).  For each of these
subslits, we determine the LE profile, calculate the offset of the
subslit to the LE using $\delta_2$ and $\Delta s$ as described in
Appendix~\ref{app:slitmisalignement}, and determine the window
function for each subslit. The total window function for the full slit
is then the flux-weighted average of the subslit window functions.

\section{Cas~A Light Echoes}
\label{sec:casa}

\subsection{Observations}

\begin{figure}
\epsscale{1.15}
\plotone{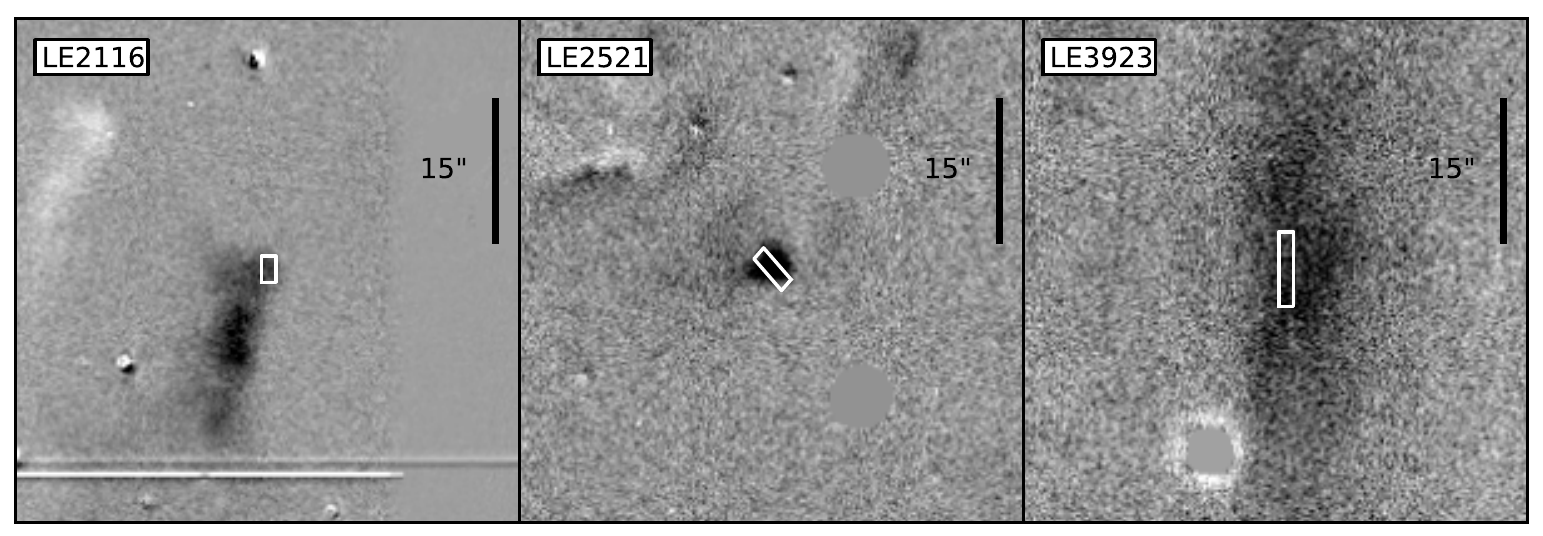}
\caption[]{Difference image cutouts for LE2116, LE2521, and LE3923
from left to right, respectively.  For all image stamps north is up
and east is left.  Excess flux in the first and second epoch is white
and black, respectively.  The position and size of the spectroscopic
slit is indicated by the white rectangle.
\label{fig:casasubstamps}}
\end{figure}

As a case study, we investigate LEs discovered in a multi-season
campaign beginning in October 2006 on the Mayall 4~m telescope at Kitt
Peak National Observatory \citep{Rest08b}.  The MOSAIC imager on the
Mayall 4~m telescope, which operates at the f/3.1 prime focus at an
effective focal ratio of f/2.9, was used with the Bernstein ``$V\!R$''
broad filter ($\lambda_{c} = 595$~nm, $\delta \lambda = 212$~nm; NOAO
Code k1040).  The imaging data was kernel- and flux-matched, aligned,
subtracted, and masked using the SMSN pipeline
\citep{Rest05a,Garg07,Miknaitis07}.  We investigate the three Cas~A SN
light echoes LE2116, LE2523, and LE3923, for which we presented spectra
in \citet{Rest10_casaspec}.  Figure~\ref{fig:casasubstamps} shows the
difference images and the slit position for these three LEs, and
Table~\ref{tab:specinfo} lists the  positions, lengths, and
position angles for each slit. LE2116 was previously reported by 
\citet{Rest08b}, whereas LE2521 and LE3923 were discovered on 
2009 September 14 and 16, respectively. For the analysis of LE2116, 
we use images from 2009 September 14, while we use the discovery images 
for LE2521 and LE3923.

\begin{deluxetable*}{cccccccccc}
\tabletypesize{\scriptsize}
\tablecaption{
\label{tab:specinfo}}
\tablehead{
\colhead{} &
\colhead{} &
\colhead{RA} &
\colhead{Dec} &
\colhead{PA SNR-LE} &
\colhead{Seeing\tablenotemark{a}} &
\colhead{Width\tablenotemark{a}} &
\colhead{Length\tablenotemark{a}} &
\colhead{PA\tablenotemark{a}} \\
\colhead{LE} &
\colhead{Telescope} &
\colhead{(J2000)} &
\colhead{(J2000)} &
\colhead{(\arcdeg)} &
\colhead{(\arcsec)} &
\colhead{(\arcsec)} &
\colhead{(\arcsec)} &
\colhead{(\arcdeg)}
}
\startdata
 LE2116 & Keck & 23:02:27.10 & +56:54:23.4 & 237.83           & 0.81 & 1.5 & 2.69 & \phantom{0}0.0  \\
 LE2521 & Keck & 23:12:03.86 & +59:34:59.3 & 299.11           & 0.79 & 1.5 & 4.32 &           41.0  \\
 LE3923 & Keck & 00:21:04.97 & +61:15:37.3 & \phantom{0}65.08 & 0.89 & 1.5 & 7.72 & \phantom{0}0.0 
\enddata
\tablenotetext{a}{Parameters for spectroscopic slit}
\end{deluxetable*}

\subsection{Cas~A Light Echo Template Spectra}
\label{sec:casa_tmplspec}

\begin{figure}
\epsscale{1.15}
\plotone{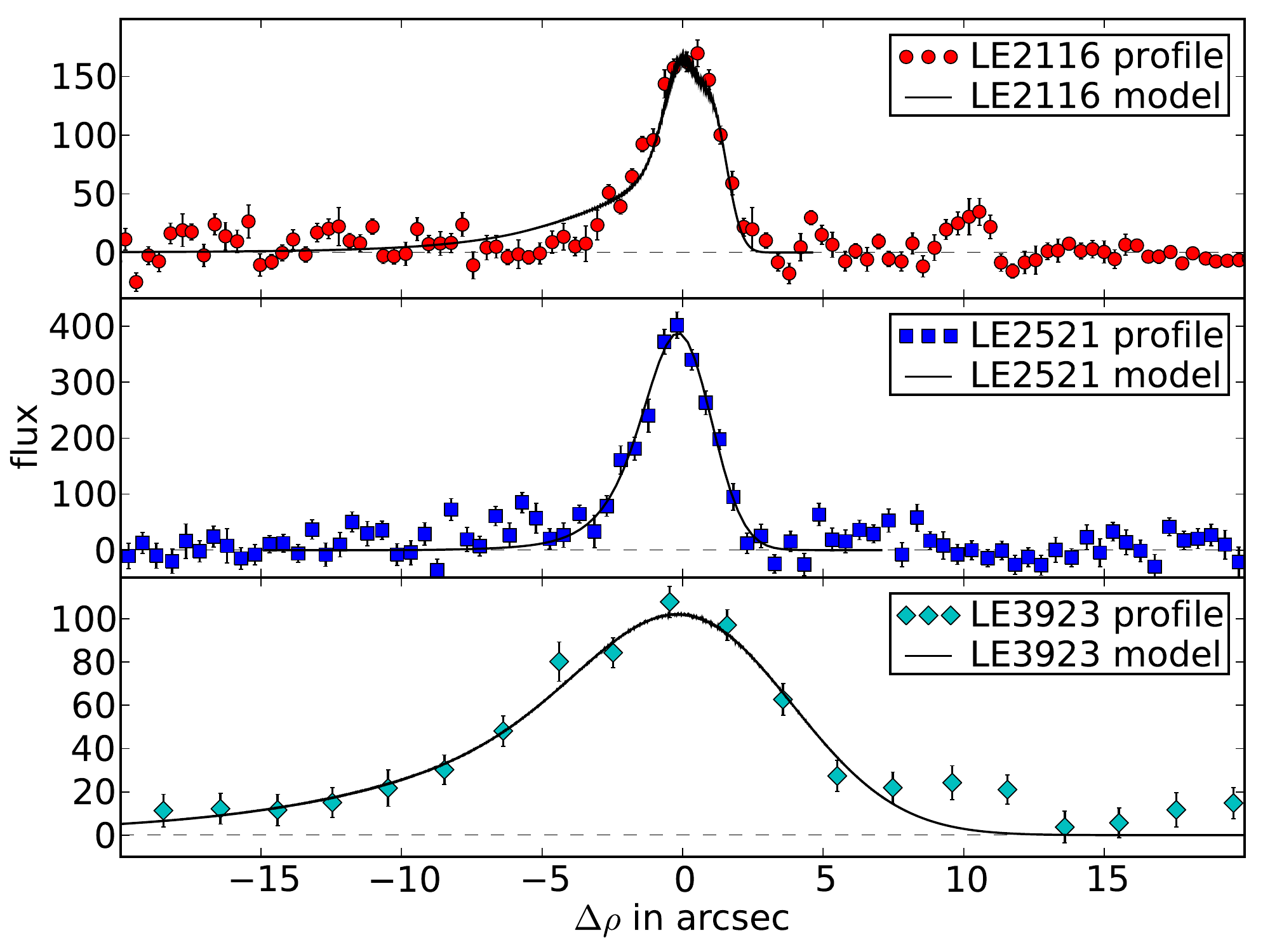}
\caption[]{LE profiles for LE2116, LE2521, and LE3923 from top
to bottom, respectively. The model fit is indicated with the black
lines.
\label{fig:casaprofiles}}
\end{figure}

There are several LEs associated with the Cas~A SN.  Here we examine
three LEs for which we have spectra.  Their profiles are shown in
Figure~\ref{fig:casaprofiles}.  To fit the LE profiles, we numerically
solve the model defined by Equation~\ref{eq:P_LE} for the observed
data.  To do this, we assume that the explosion occurred in the year $1681
\pm 19$ \citep{Fesen06_expansionasym_age}, determine the scattering
dust coordinates $\rho_{0}$ and $z_{0}$ by using Equation~\ref{eq:le}
with the measured angular separation between the LE and the SNR, and
we determine the dust inclination from the apparent motion of the LEs
(this process is defined in Appendix~\ref{app:inclination} and
described in more detail by Rest et~al., in prep.).  The remaining
free model parameters are the LE profile peak height and position and
the dust filament width, which has physical implications.  For the
three LEs presented above, we display their best fit model profiles in
Figure~\ref{fig:casaprofiles}.

\begin{figure}
\epsscale{1.15}
\plotone{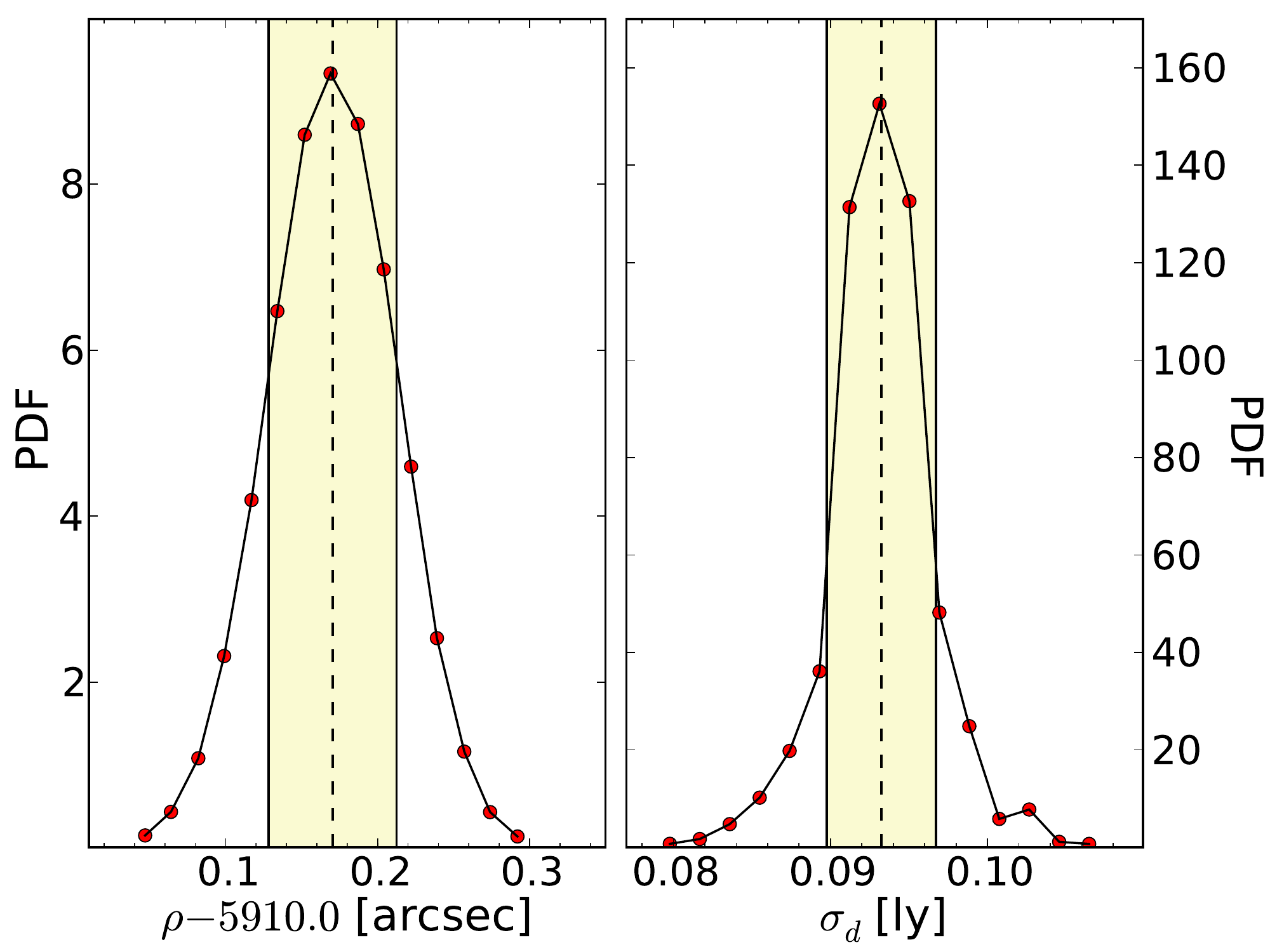}
\caption[]{Probability density functions (PDFs) for the peak position
(left) and dust width (right) of LE2521 at $u = 0.5\arcsec$.  The most
likely value is indicated with a dashed line, and the yellow shaded
region indicates the 1-$\sigma$ interval.
\label{fig:marginalization}}
\end{figure}

To numerically fit the model parameters, we determine the best fitting
profile by calculating the reduced $\chi^{2}$ for a 3-dimensional grid
of the three fit parameters.  For each parameter, we marginalize over
the other parameters to obtain the best-fitting value.  The initial
step sizes of the parameter grid are often too large for a proper
marginalization.  To ensure a proper determination of the parameters,
we iterate this process three times adjusting the grid spacing, guided
by the fitted parameters and their uncertainties from the previous
iteration.  In all but the lowest signal-to-noise ratio cases, the
probability density functions have a well-behaved Gaussian shape.
Figure~\ref{fig:marginalization} shows as examples the marginal
probability density functions (PDFs) for the peak position and dust
width of the LE2521 LE profile at $u = 0.5\arcsec$.

Table~\ref{tab:casaprofiles} shows the input and fitted parameters of
these three LE profiles.  These profiles are for subslits with $u =
0.5\arcsec$, i.e., close to the center of the spectroscopic slit.  The
definition of the ($\rho, u$) coordinate system and how it relates to
the LE and the subslits is explained in detail in
Section~\ref{sec:slitoffset}, Appendix~\ref{app:slitmisalignement} and
Figure~\ref{fig:slitpos_illu}.  LE2116 and LE2521 have profiles with a
similar width.  The profile of LE3923, however, has a significantly
larger width. Note that the model fits the data very well.

\begin{deluxetable*}{cccccccccccc}
\tabletypesize{\scriptsize}
\tablecaption{
\label{tab:casaprofiles}}
\tablehead{
\colhead{} &
\colhead{Image} &
\colhead{Seeing} &
\colhead{$\alpha$} &
\colhead{$\delta_1$} &
\colhead{$\delta_2$} &
\colhead{$\Delta \rho_{\rm offset}$} &
\colhead{$\rho_{0}$} &
\colhead{$z_{0}$} &
\colhead{$\sigma_{d}$} &
\colhead{$S\!B$} &
\colhead{$\sigma_{s, {\rm eff}}$} \\
\colhead{LE} &
\colhead{UT Date} &
\colhead{(\arcsec)} &
\colhead{(\arcdeg)} &
\colhead{(\arcdeg)} &
\colhead{(\arcdeg)} &
\colhead{(\arcsec)} &
\colhead{(ly)} &
\colhead{(ly)} &
\colhead{(ly)} &
\colhead{(mag)} &
\colhead{(\arcsec)}
}
\startdata
 LE2116 & 20090914 & 0.97 & \phantom{0}9(5)  & \phantom{-}\phantom{0}7.80 & $-39.97$           & $-0.42$ & \phantom{0}627.9561(18) & \phantom{0}437.11             & 0.031(10) & 22.82(02) & 1.77 \\
 LE2521 & 20090914 & 1.13 & 54(5)            & $-10.98$                   & \phantom{0}$-0.91$ & $-0.01$ & \phantom{0}317.9613(23) & \phantom{0}\phantom{0}$-9.89$ & 0.093(04) & 21.83(02) & 1.53 \\
 LE3923 & 20090916 & 1.65 & \phantom{0}7(5)  & \phantom{-}10.44           & $-35.36$           & $-0.35$ & 1203.8904(76)           & 2045.38                       & 0.614(53) & 23.35(04) & 1.65
\enddata
\tablecomments{
All the values are for the subslits with $u=0.5$. 
}
\end{deluxetable*}

\begin{figure}
\epsscale{1.15}
\plotone{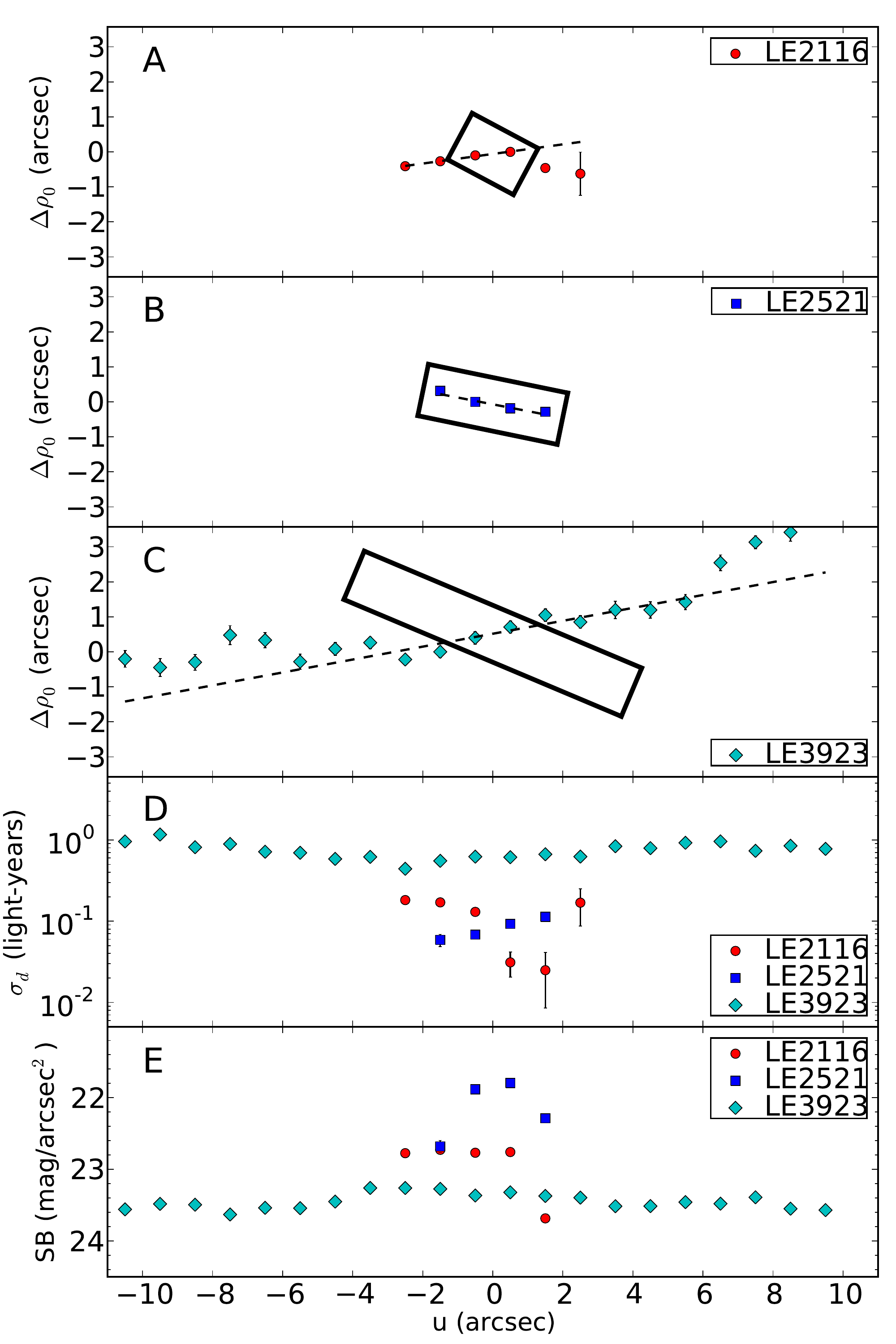}
\caption[]{Panel \emph{A}, \emph{B}, and \emph{C} show the positions
of the LE profile peaks for the different subslits for LE2116, LE2521,
and LE3923, respectively. $\Delta\rho_{0}(u)$ is defined as
$\rho_{0}(u)-\rho_{0}(0)$, and $u$ is orthogonal to the $\rho$-axis
through $\rho_{0}(0)$. Red, blue, and cyan symbols indicate LE2116,
LE2521, and LE3923, respectively.  Panel \emph{D} shows the fitted
dust widths, $\sigma_{d}$, and Panel \emph{E} shows the surface
brightnesses.
\label{fig:casafitparams}}
\end{figure}

Figure~\ref{fig:casafitparams} shows the fitted LE peak position
$\rho_{0}$, dust width $\sigma_{d}$, and peak surface brightness
$S\!B$ for all subslits of each of the three LEs.  Panels~A--C show
$\rho_{0}$ for the three LEs, with the spectroscopic slit overplotted
(black solid lines). The dashed line indicates a straight line fit to
the subslits that are within the spectroscopic slit.  Even though
there is some structure in the $\rho_{0}$ versus $u$ relation, locally
the LE position can be approximated very well with a straight line.
Panel~D shows the dust width $\sigma_{d}$.  The dust width of LE3923
is nearly constant over $20\arcsec$ and is an order of magnitude
larger than the dust width of the other two LEs.  The dust width for
LE2116 decreases and then increases over only \about $2\arcsec$.  The
width decreases to $\sigma_{d} = 0.03$~ly for $u \ge 0.5$, making
this portion of the LE2116 filament the thinnest yet observed.
We discuss potential scientific opportunities for a LE with such a
thin dust width in Section~\ref{sec:casa_constraininglc}.  Panel~E
shows the surface brightness.  
LE2521 has a surface brightness brighter than 21 mag/arcsec$^{-2}$
and is the brightest Cas~A LE yet reported.
However, it is spatially small.  On the other
hand, LE3923 is relatively faint, but its large size makes
spectroscopy productive.

For the spectra presented by \citet{Rest10_casaspec}, the profile fits
had not been completed when slit masks were designed.  As a result,
the alignment between the LEs and the slit were not optimal.  In
particular, there is a large misalignment in angle for LE3923;
however, that LE has a large profile (${\rm FWHM} \approx 10\arcsec$;
see bottom panel of Figure~\ref{fig:casaprofiles}) and therefore an
offset at the 1--2$\arcsec$ scale does not decrease the total amount
of flux by a significant amount.  However, it does affect the observed
spectra since the window function and effective light curves strongly
depend on these offsets.

For each LE, we calculate the window function by fitting the observed
LE profile with the numerical model as described above.  The
spectroscopic slit is applied to the fitted LE profile using
Equations~\ref{eq:rhomin}--\ref{eq:Hw} as described in
Appendix~\ref{app:lespec}.  It is important to note that now the
seeing from the spectroscopic observation is used instead of the
seeing from the image used to determine the LE profile.  For each
slit, the weighted average of all subslits is calculated using
Equations~\ref{eq:delta_rho_offset} -- \ref{eq:wslit} in
Appendix~\ref{app:slitmisalignement}.

\begin{figure}
\epsscale{2.3}
\plottwo{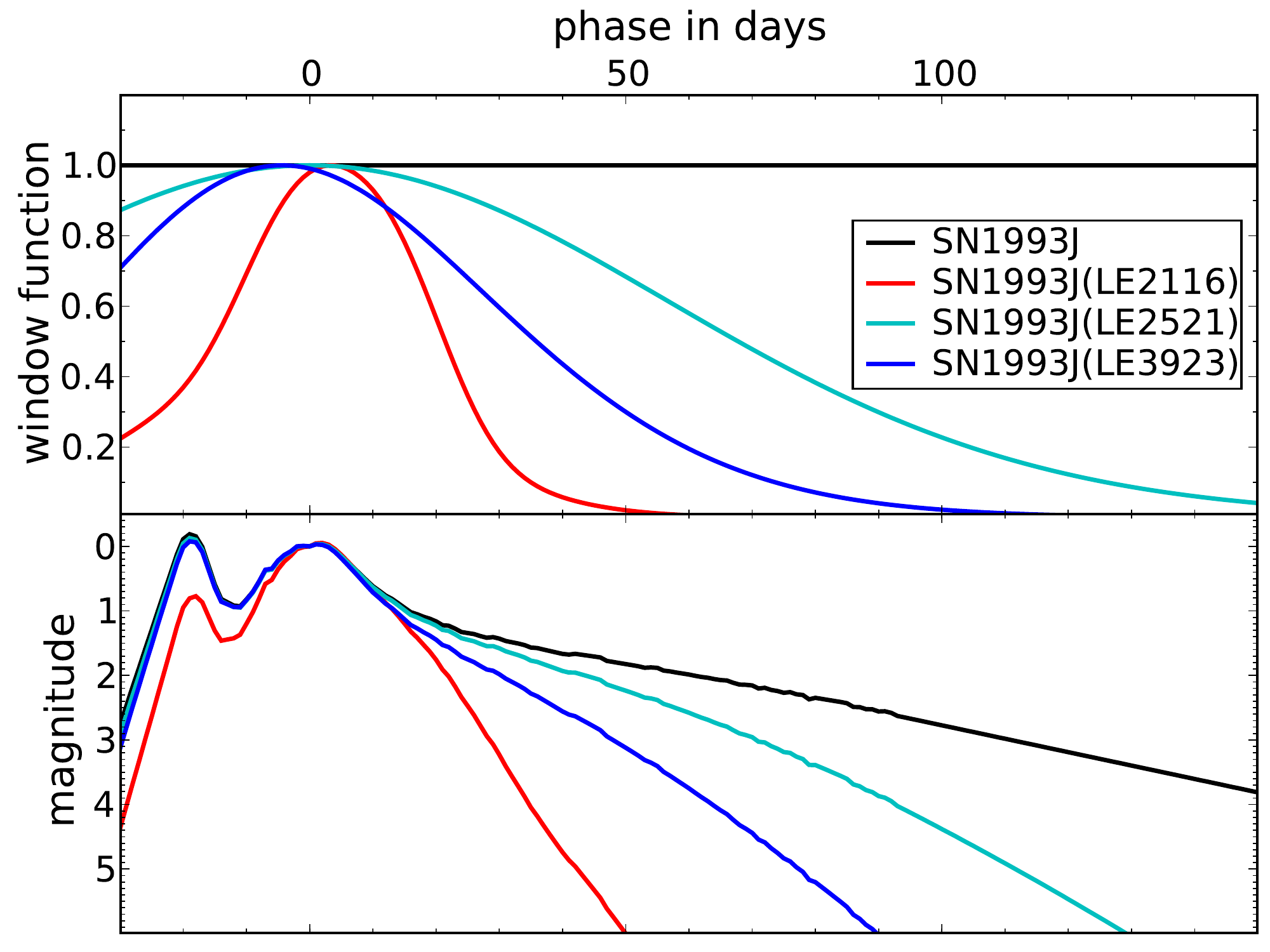}{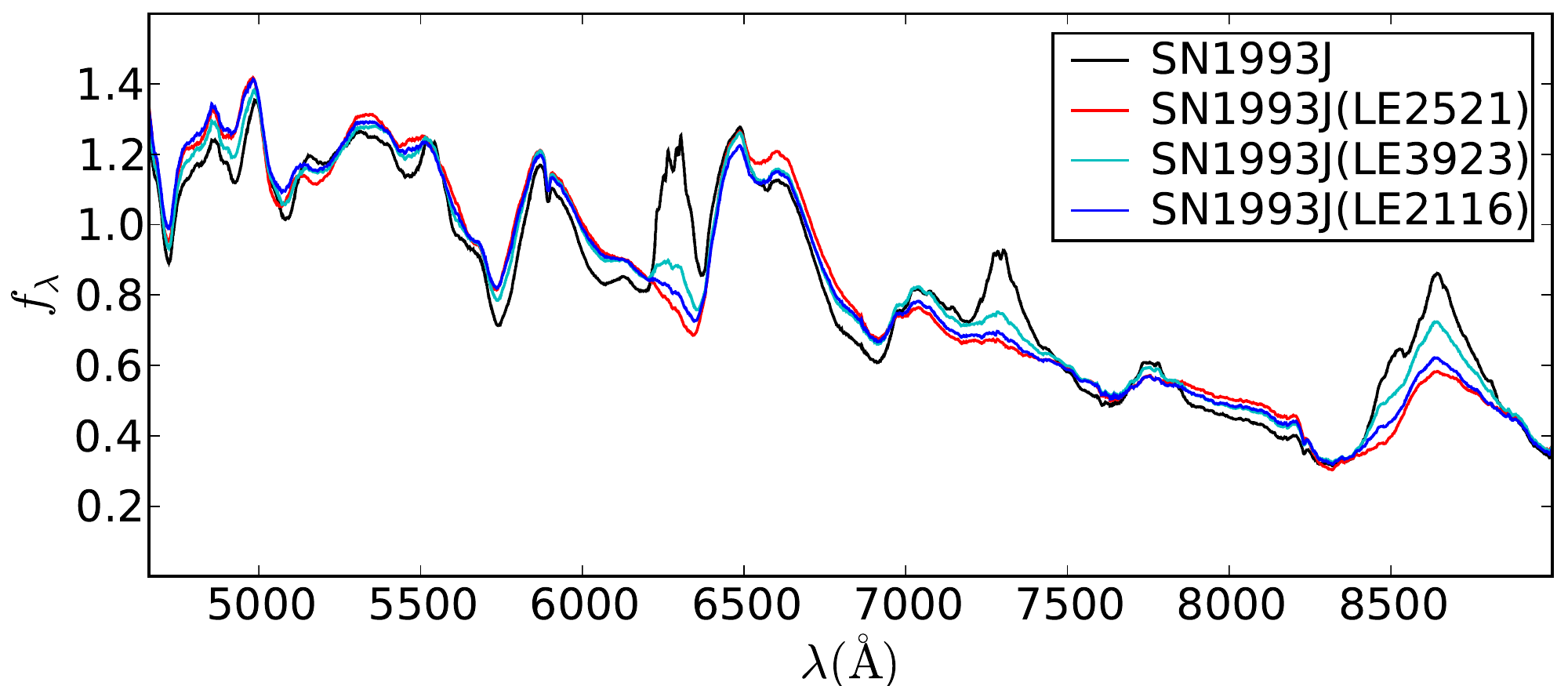}
\caption[]{\emph{Top:} window function for the
different LEs.  \emph{Middle:} effective light curves of SN~1993J for
the LEs.  The unmodified light curve of SN~1993J is shown in
black. \emph{Bottom:} integrated SN~1993J spectra, where the
integration is weighted by the respective effective light curve.  Red,
cyan, and blue curves represent LE2116, LE2521, and LE3923,
respectively.  The black spectrum is weighted by the original,
unmodified SN~1993J light curve.  The differences in the spectra are
mainly due to the strong [\ion{O}{1}] $\lambda\lambda 6300$, 6363,
[\ion{Ca}{2}] $\lambda\lambda 7291$, 7324, and \ion{Ca}{2} NIR triplet
features in the late-phase spectra.
\label{fig:casawf}}
\end{figure}

The top panel of Figure~\ref{fig:casawf} shows the window functions
for the three LEs for which we have spectra.  The middle panel shows
the SN~1993J light curve and the effective light curves for each LE
(using the LE profiles and assuming that their light curves are the
same as SN~1993J).

Even though LE2521 and LE2116 have comparable dust widths, the dust
inclination for LE2521 is significantly larger, resulting in a wider
window function than LE2116.  This also causes the projected SN
light-curve shape to be ``squashed.''  LE3923 has a similar
inclination to that of LE2116, but its scattering dust is thicker,
also leading to a wider window function. Note that the three window
functions, and therefore also the effective light curves (see middle
panel in Figure~\ref{fig:casawf}), have significantly dropped off
after 100~days past maximum.  However, the decline time of SN is
significantly longer, and consequently the contribution of spectra
with phases $\ge 100$~days is cut off - a fact illustrated in the
bottom panel of Figure~\ref{fig:casawf}, which shows the integrated
spectra for SN~1993J. The late type spectra of SN~IIb like SN~1993J
are dominated by [\ion{O}{1}] $\lambda\lambda 6300$, 6363,
[\ion{Ca}{2}] $\lambda\lambda 7291$, 7324, and the \ion{Ca}{2} NIR
triplet. Note that the black spectrum, which is weighted by the
original, un-modified light curve, is much stronger at the position 
of these lines compared to the other spectra, which are integrated using the
effective light curves. We conclude that a perfectly symmetrical 
SN observed at different position angles might be erroneously interpreted 
as asymmetrical if scattering dust properties, seeing, and slit width 
are not taken into account.

\subsection{Light Echo Spectra: Systematics}
\label{sec:systematics}

We have shown in Section~\ref{sec:analysis} with simulated light
curves and dust filaments how the LE spectra are influenced by the
dust properties and observing conditions, and in
Section~\ref{sec:casa_tmplspec} we have applied this to the real-world
example of the Cas~A LEs. In this section, we investigate how changing
these properties impacts the LE spectra.  We start with the fit to the
Cas~A LE profiles of LE2521 described in
Section~\ref{sec:casa_tmplspec}, and create a window function,
effective light curve, and LE spectrum using using the spectroscopic
and photometric library of SN~1993J as a template. We then vary dust
width, dust inclination, and seeing while keeping all other parameters
fixed.

\begin{figure}
\epsscale{1.2}
\plotone{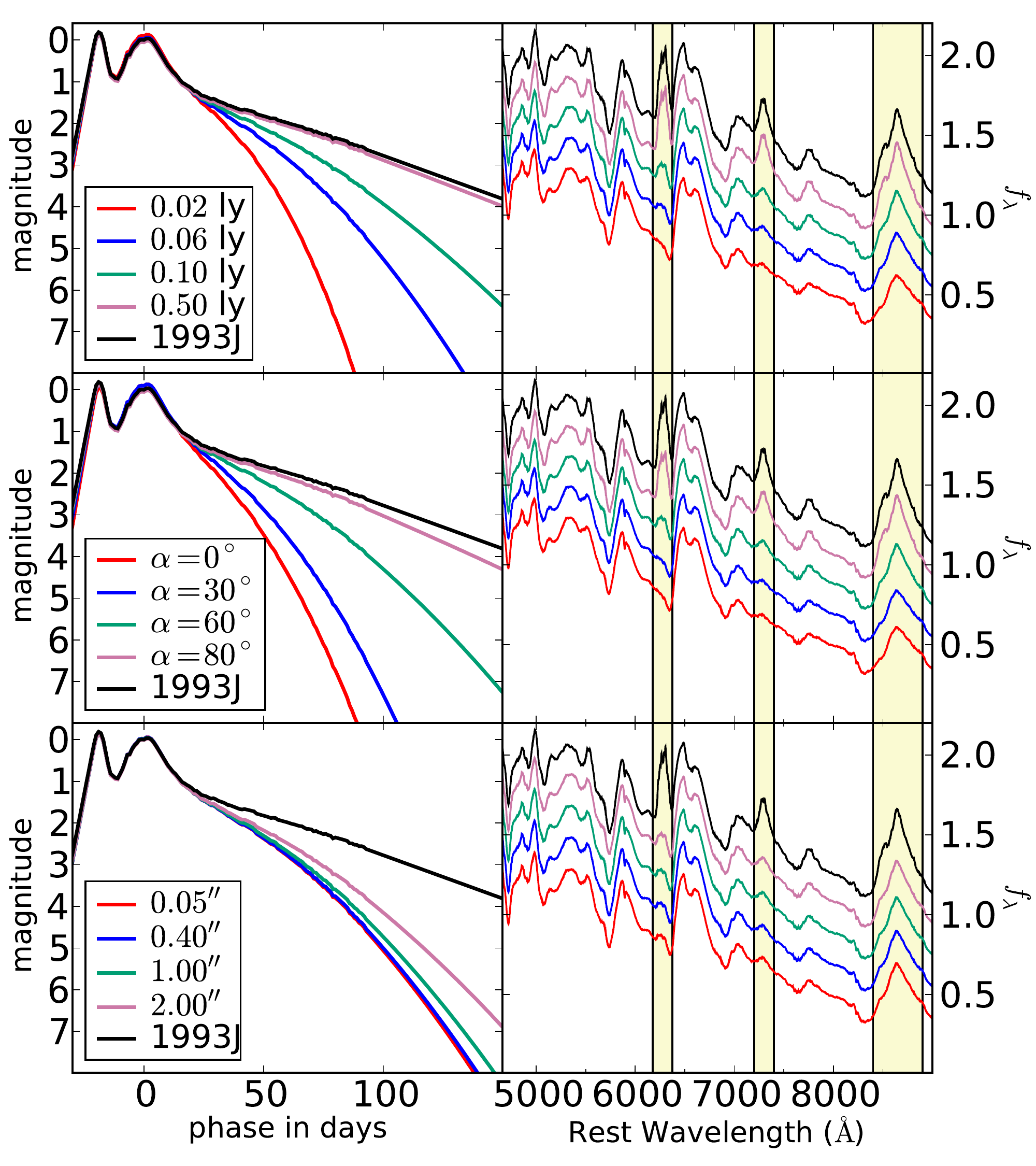}
\caption[]{Effective light curves and integrated SN~1993J spectra
varying the dust width, $\sigma_{d}$ (top row), the dust inclination,
$\alpha$ (middle row), and the seeing (bottom row) for LE2521.  For a
given row, only one parameter varies, with the other parameters are
set by the fit described above and shown in
Figure~\ref{fig:casaprofiles}. The black line indicates the original
SN~1993J light curve and the corresponding integrated spectra.
\label{fig:casa_changeparams}}
\end{figure}

Figure~\ref{fig:casa_changeparams} shows the effective light curves
(left panels) and integrated SN~1993J spectra (right panels).  In the
top panel, we vary the dust width, $\sigma_{d}$.  We find that,
consistent with the simulations in Section~\ref{sec:analysis}, the
nebular [\ion{O}{1}] and [\ion{Ca}{2}] lines in the integrated spectra
have a strong dependence on the dust width. For $\sigma_{d} \lta
0.1$~ly, the [\ion{O}{1}] and [\ion{Ca}{2}] emission lines are weak,
whereas for larger values the integrated spectrum quickly approaches
the limit of the integrated spectrum without any window function
applied with strong [\ion{O}{1}] and [\ion{Ca}{2}] emission lines.  We
find similar results for the dust inclination: the difference between
the integrated spectra is small and the [\ion{O}{1}] and [\ion{Ca}{2}]
lines are weak for $\alpha \lta 60$\arcdeg, but then rapidly
approaches the integrated spectrum without any window function applied
for $\alpha$ close to 90\arcdeg. We find that the seeing (bottom
panel) only has a very strong impact on the observed LE spectrum if
the dust width is very small as discussed in
Section~\ref{sec:casa_timeresolve} and \ref{sec:casa_constraininglc}.

\begin{figure}
\epsscale{1.2}
\plotone{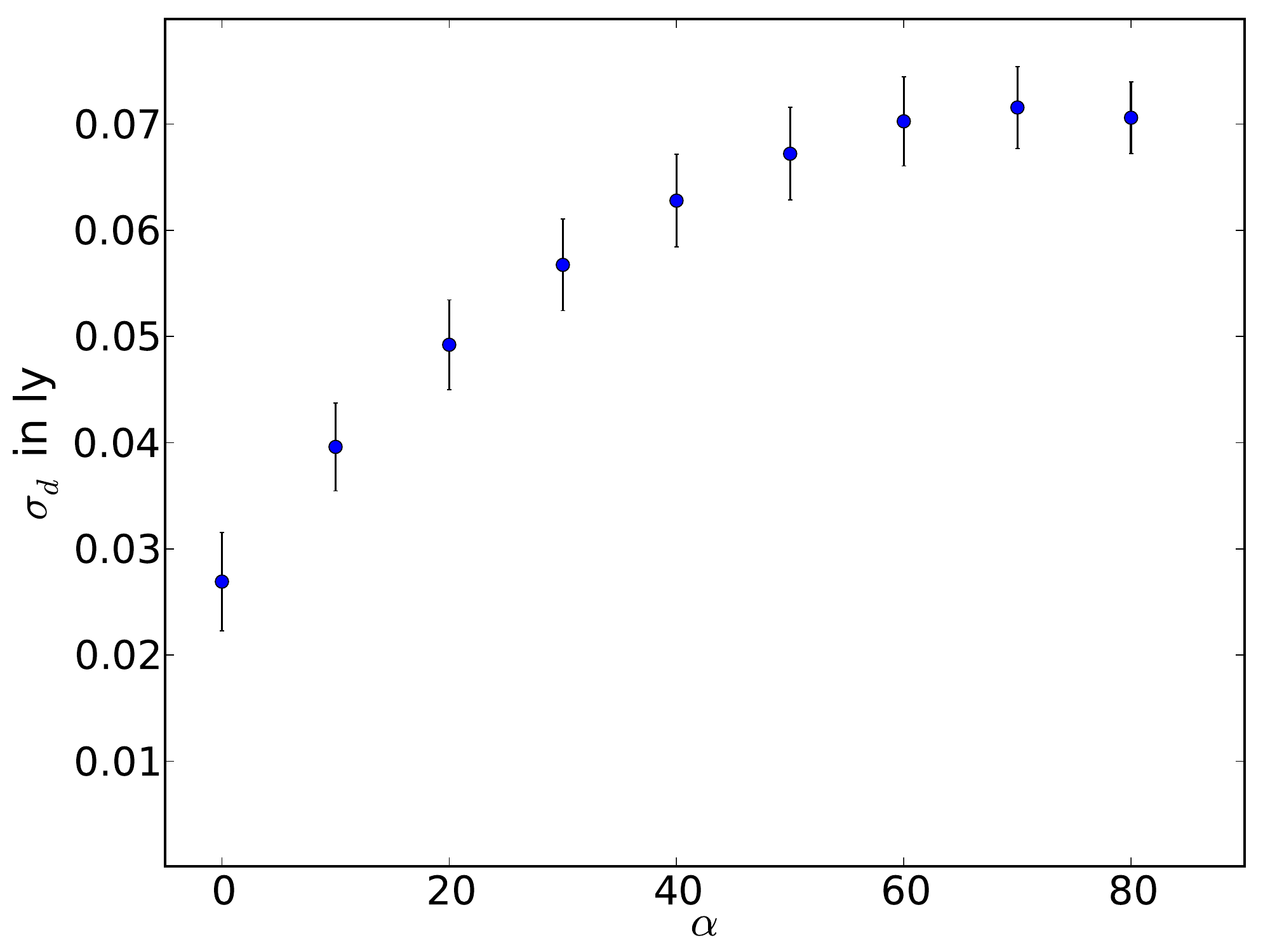}
\caption[]{
Fitted dust width, $\sigma_{d}$, for the LE2521 LE profile for
different input dust inclination, $\alpha$.
\label{fig:casa_varalpha_refit_dustwidth}}
\end{figure}

\begin{figure}
\epsscale{2.3}
\plottwo{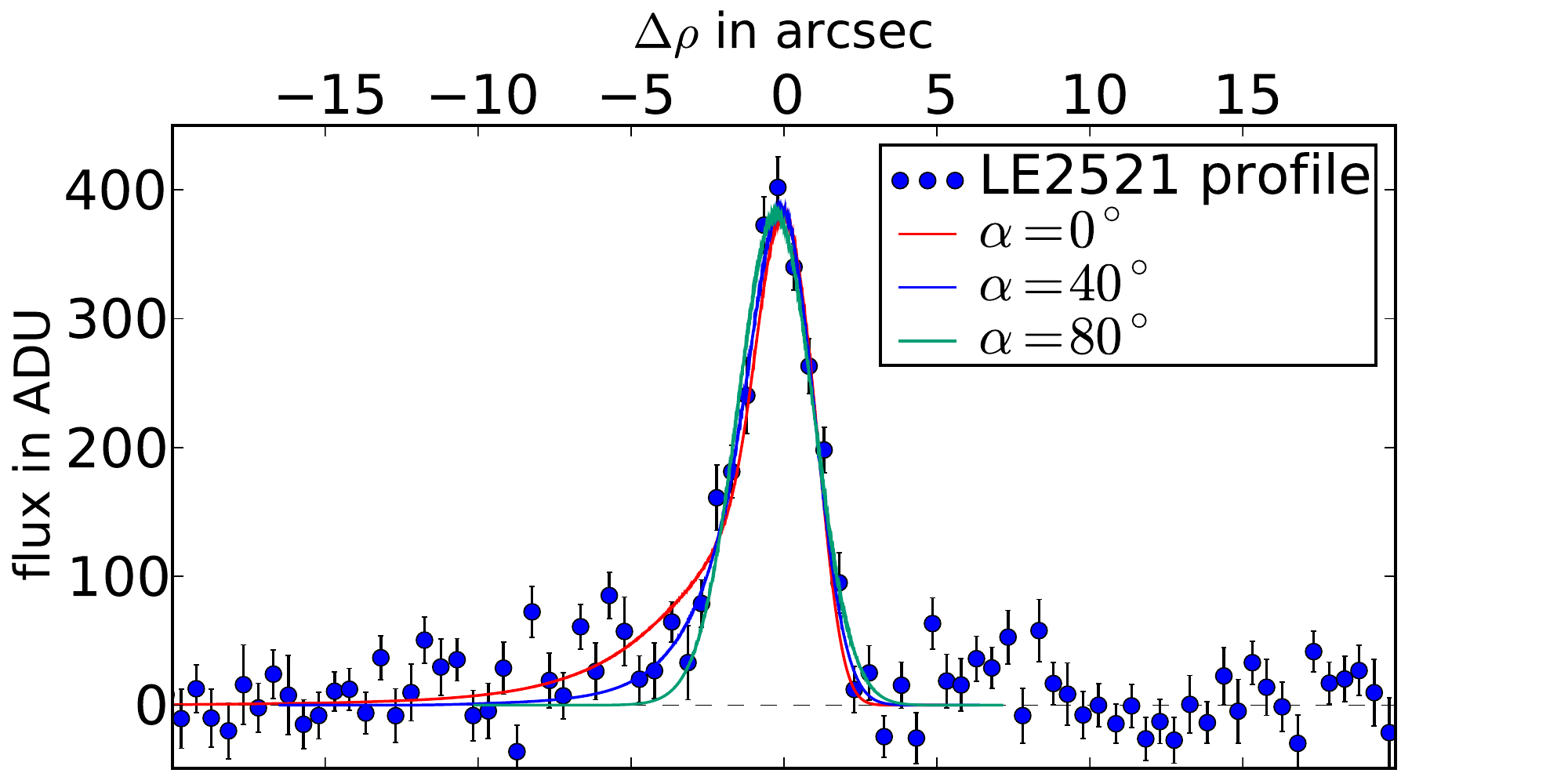}{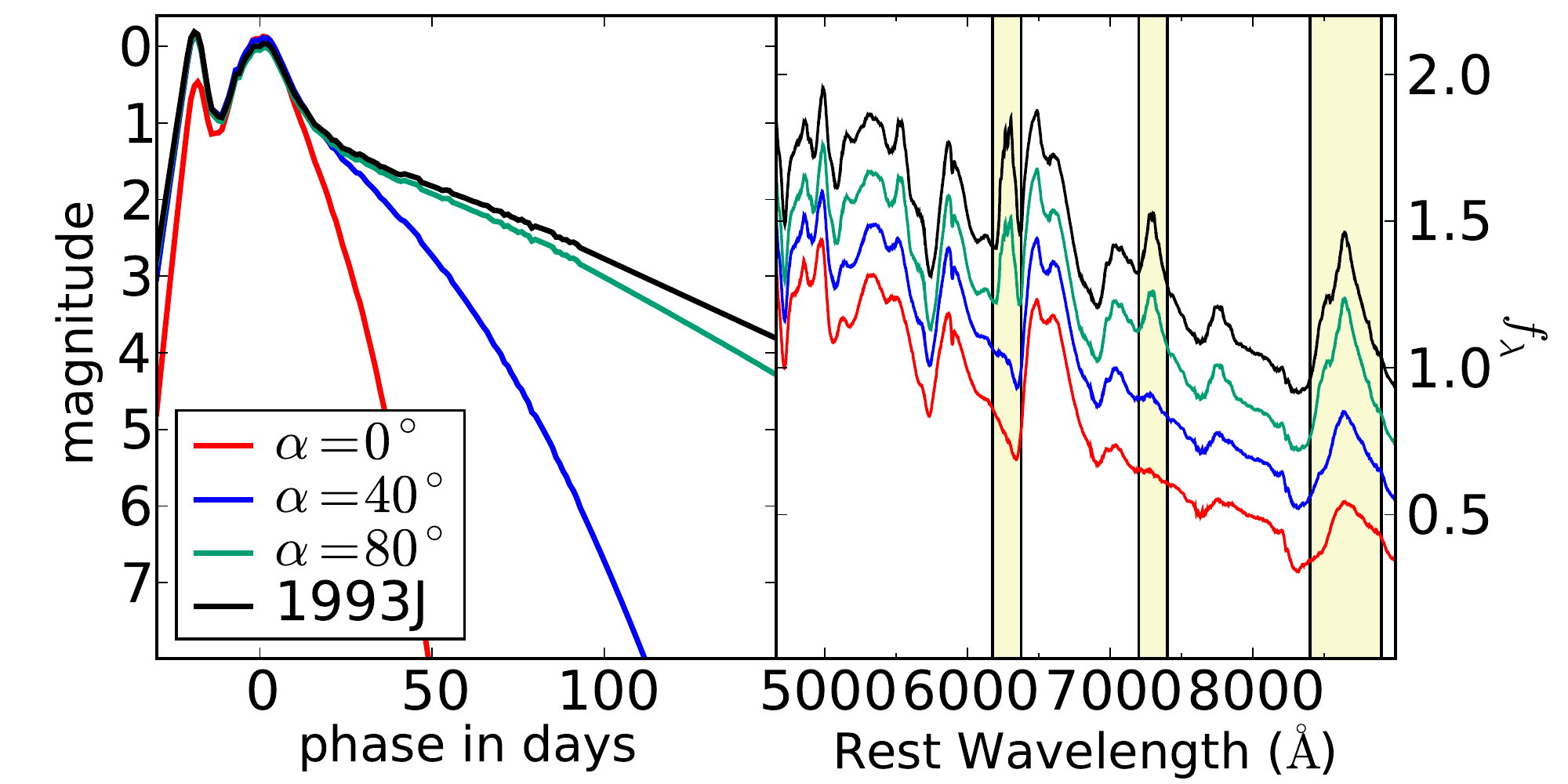}
\caption[]{
\emph{Top:} LE2521 profile and model fits for varying inclination
angles, $\alpha$.  \emph{Bottom:} Effective light curves (left) and
the corresponding integrated spectra (right).  The black line
indicates the original SN~1993J light curve and the corresponding
integrated spectra.
\label{fig:casa_varalpha_refit}}
\end{figure}

We also investigate how uncertainties in the dust inclination impact
the derived integrated spectra template.  To do this, we fixed the
dust inclination from 0\arcdeg\ to 80\arcdeg for LE2521, and fit the
LE profile.  Figure~\ref{fig:casa_varalpha_refit_dustwidth} shows the
fitted dust width, $\sigma_{d}$, with respect to the input dust
inclination, $\alpha$.  Three examples of the fitted LE profiles are
shown in the top panel of Figure~\ref{fig:casa_varalpha_refit}. The
bottom panels show the corresponding effective light curves (left) and
integrated spectrum (right).

Note that while the LE profile fit is very good in all cases, the
effective light curves are different, and thus the integrated spectra
differ significantly in [\ion{O}{1}] and [\ion{Ca}{2}] line strengths.
The reason for these differences is the degeneracy between the dust
inclination and the dust width.  For decreasing dust inclination
(i.e., the dust inclination gets closer to the tangential of the LE
paraboloid), the projected light curve is ``stretched.''  This effect
can be nullified by decreasing the dust width, resulting in a similar
fit to the LE profile with significantly different parameters (see
Figure~\ref{fig:casa_varalpha_refit_dustwidth}).

For inclinations close to 90\arcdeg, the LE profile is dominated by
the dust profile and not by the projected light-curve profile.  Such a
situation manifests itself in a LE profile that is symmetrical (see
green line in top panel of Figure~\ref{fig:casa_varalpha_refit}) in
contrast to the asymmetrical LE profile shape (see red line in top
panel of Figure~\ref{fig:casa_varalpha_refit}) that still reveals
indications of the original light curve.  Similarly, the fitted
$\sigma_{d}$ is essentially constant for $\alpha \approx 90\arcdeg$.

The tests in this section have shown that an incorrect dust
inclination can systematically bias the interpretation of SN
properties from LE spectra.  We discuss the sources of uncertainty in
the determination of the inclination in
Appendix~\ref{app:inclination}.

\section{SN~1987A Light Echoes}
\label{sec:87A}

\subsection{Observations}

Beginning in 2001, the \mbox{SuperMACHO} Project microlensing survey
employed the CTIO 4~m Blanco telescope with its 8K~$\times$~8K MOSAIC
imager (plus its atmospheric dispersion corrector) to monitor the
central portion of the LMC every other night for 5 seasons (September
through December).  The images were taken through our custom ``\VR''
filter ($\lambda_{c} = 625$~nm, $\delta \lambda = 220$~nm; NOAO Code
c6027).  SN~1987A and its LEs are in field sm77, which was
observed with exposure times between 150 and 200 seconds.  We have
continued monitoring this field since the survey ended. 
The difference imaging reduction process was identical to
the KPNO Cas~A imaging.

\subsection{Profile Modeling}
\label{sec:87a_model}

The ring-shaped LEs associated with SN~1987A have previously been
studied in great detail, mapping the dust structure surrounding the SN
\citep[e.g.,][]{Sugerman05b}.  Unlike historical SNe, a complete
spectral and photometric history of the SN is available for SN~1987A.
This historical database allows an observed LE spectrum to be modeled
unambiguously against a spherically symmetric model for the explosion.
In addition, the high signal-to-noise ratio of the SN~1987A LEs make
them ideal for our tests.

\begin{figure}
\epsscale{1.15}
\plotone{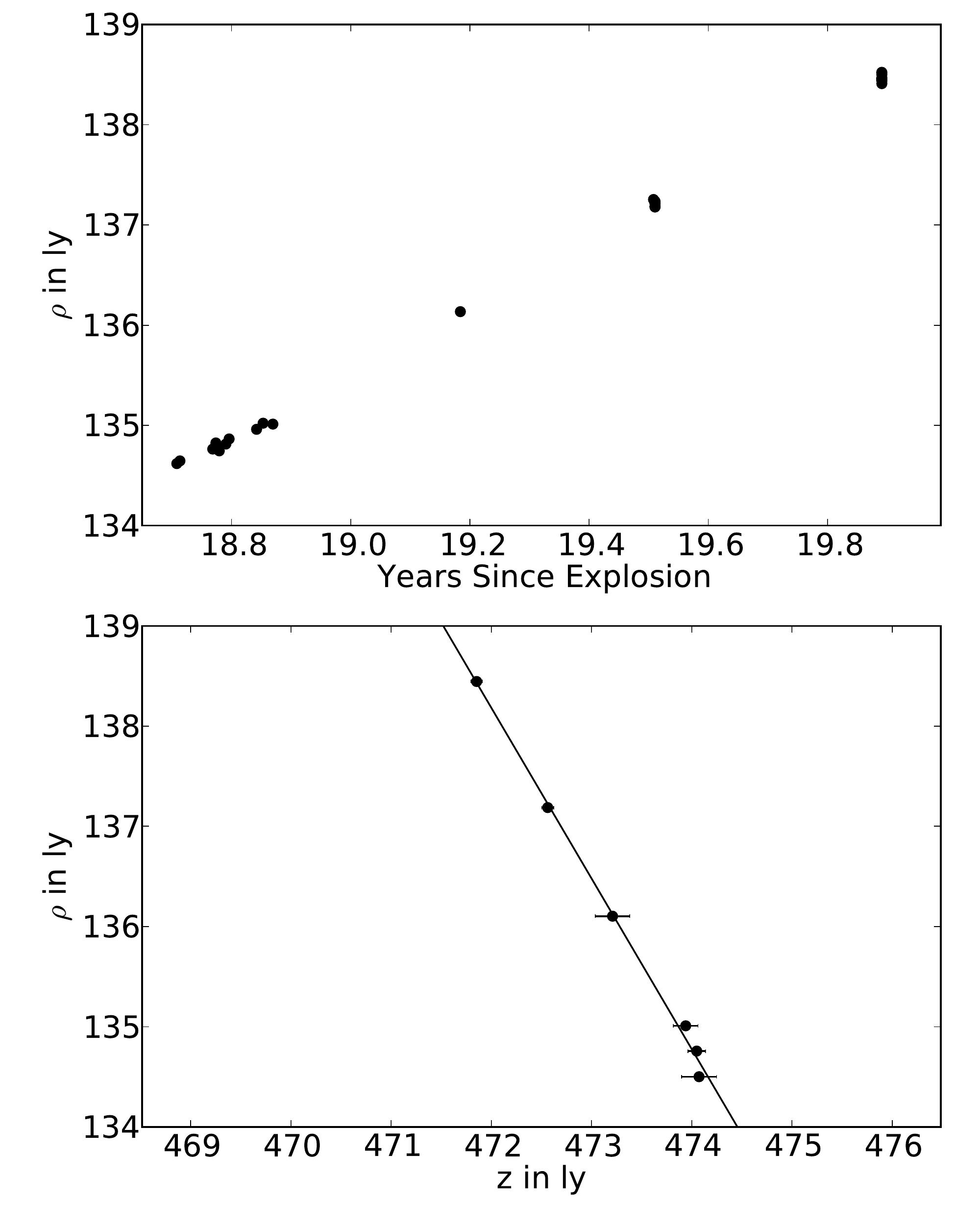}
\caption[]{\emph{Top:} Apparent motion of a single position
angle of the LE system of SN~1987A.  The uniform motion of the LE on
the sky indicates a unique inclination for the dust structure.  The
superluminal motion of the LE is $4.09 \pm 0.02~\arcsec$/year.
\emph{Bottom:} The above $(\rho,t)$ data is binned when observations
are taken within 10 days of each other, and converted to the $(z,
\rho)$ plane using equation~\ref{eq:le}.
The resulting fit (solid line) gives an inclination of $\alpha = 30.4
\pm 0.8$\arcdeg.
\label{fig:87a_appmot_and_inc}}
\end{figure}

We consider an example LE from SN~1987A corresponding to a unique
position angle on the sky.  The spectrum associated with this LE was
taken using the Gemini Multi-Object Spectrograph
\citep[GMOS;][]{Hook04} on Gemini-South, using the R400 grating.  The
spectroscopic slit was $1.0\arcsec$ wide and $2.16\arcsec$ long, with
the slit tilted to be tangential to the observed LE ring.  The upper
panel of Figure~\ref{fig:87a_appmot_and_inc} shows the apparent motion
of the LE monitored over more than a year. The motion on the sky is
constant on such time scales ($4.09 \pm 0.02$~\arcsec/year),
indicating constant inclination of the dust structure.  To determine
the angle of inclination, $\alpha$, we translate ($\rho, t$) into $(z,
\rho)$ as described in Appendix~\ref{app:inclination}, resulting in a
dust inclination of $\alpha = 30.4 \pm 0.8$\arcdeg\ (see lower panel
of Figure~\ref{fig:87a_appmot_and_inc}).

\begin{figure*}
\epsscale{1.15}
\plotone{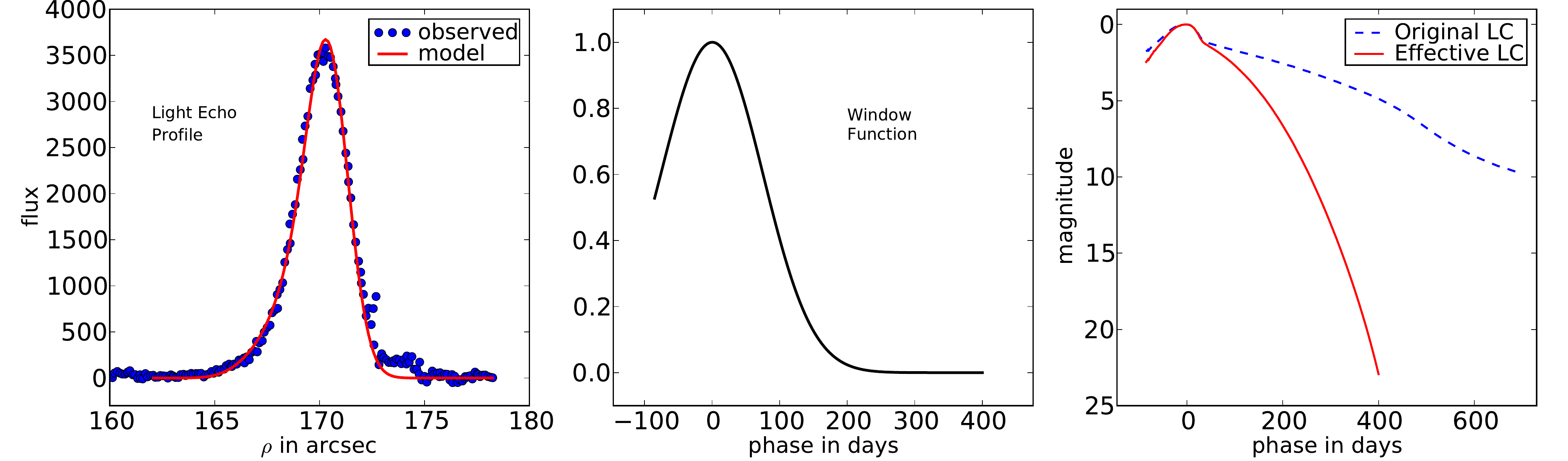}
\caption[]{\emph{Left:} observed profile for the SN~1987A LE, with the
best-fit model profile over-plotted.  \emph{Middle:} resulting window
function associated with the $1.0\arcsec$ wide spectroscopic slit.
\emph{Right:} effective light curve from the model compared with the
original light curve of SN~1987A.
\label{fig:87a_example}}
\end{figure*}

A high signal-to-noise ratio difference image is used to model the
effect of the dust on the spectroscopic slit. Input parameters are the
inclination ($\alpha = 30.4$\arcdeg), observed seeing for the image
($1.22\arcsec$), and delay time since explosion ($ct =
19.9$~years). The photometric data of SN~1987A \citep{Hamuy88,
Suntzeff88, Hamuy90} are used to model the LE profile observed on the
sky, with the best-fitting model shown in the left panel of
Figure~\ref{fig:87a_example}.  The dust structure model is able to
recreate the observed LE profile exceedingly well. The window function
and resulting effective light curve associated with the spectroscopic
slit (convolved to the seeing at time of spectroscopy) are shown in
the middle and right panels of Figure~\ref{fig:87a_example},
respectively. The average dust width associated with the spectroscopic
slit is $\sigma_{d} = 4.26 \pm 0.06$~ly ($1.31 \pm 0.02$~pc).  We note
that at the inclination and dust widths observed here, the model is
less sensitive to uncertainties in $\alpha$.  Testing of the model
shows that at such inclinations, uncertainties in $\alpha$ on the
order of 5\arcdeg\ have little impact on the window function and
resulting integrated spectrum (see Section~\ref{sec:alpha_changes}).

\begin{figure}
\epsscale{1.2}
\plotone{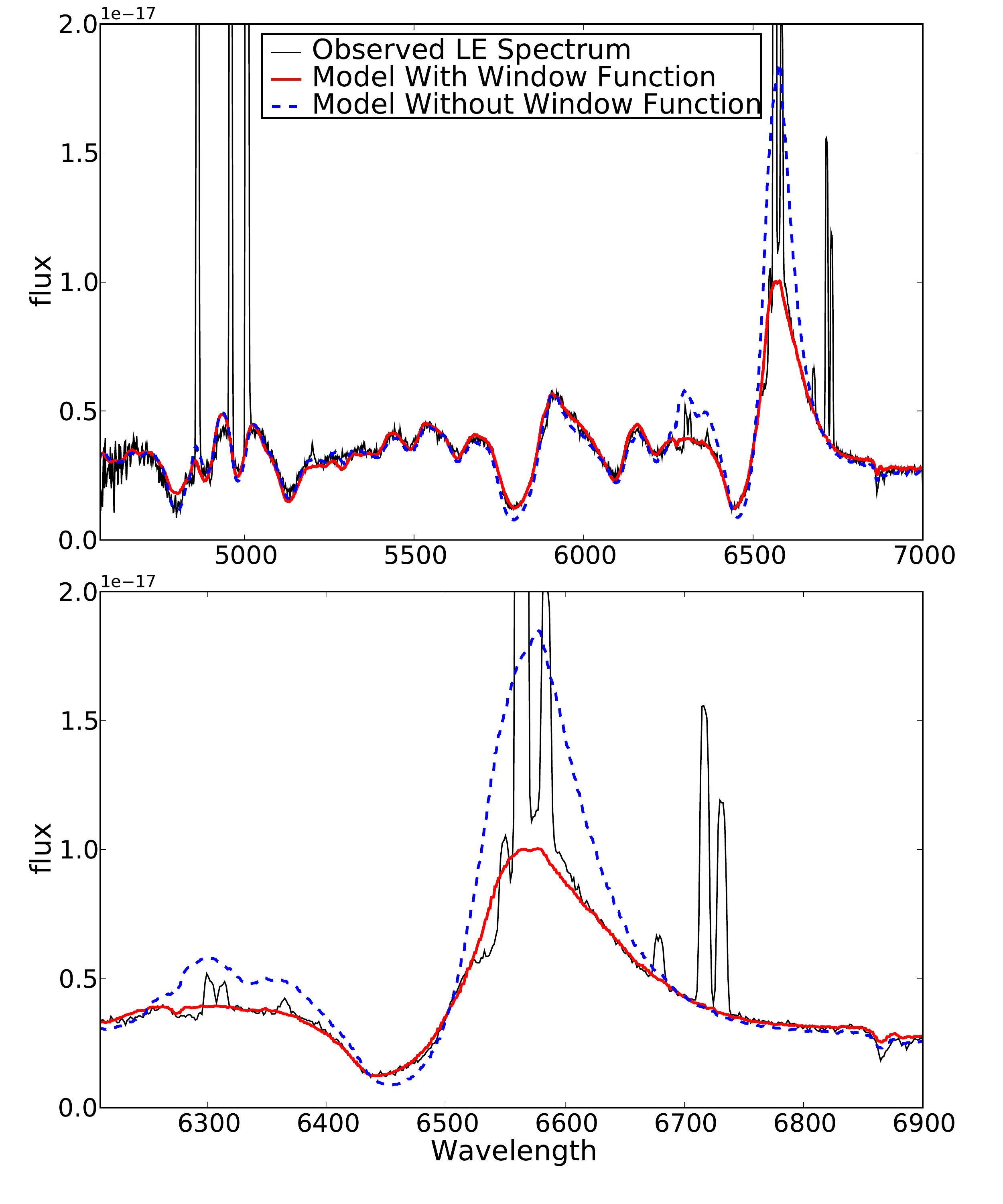}
\caption[]{\emph{Top:} Observed LE spectrum of SN~1987A (black) with
the best-fit model (using the window function from
Figure~\ref{fig:87a_example}) integrated spectrum (red) and the full
light-curve weighted integrated spectrum (blue).  \emph{Bottom:}
Detail of the spectra near the H$\alpha$ feature.  Note that the
narrow lines dominating the emission peak of the profile are strong
nebular emission lines, and are not associated with the SN.  The
best-fit model integrated spectrum fits the observed LE spectrum well
in this region, while the full light-curve weighted integrated
spectrum is a poor fit.
\label{fig:87a_specfit}}
\end{figure}

The impact of the scattering dust is clear when fitting the model to
the observed spectrum.  To integrate the spectral history of SN~1987A
we take advantage of the extensive spectral database from SAAO and CTIO
observations after the explosion \citep{Menzies87, Catchpole87,
Catchpole88, Whitelock88, Catchpole89, Whitelock89,CTIO_87aspec1,CTIO_87aspec2}.
Figure~\ref{fig:87a_specfit} shows the observed LE spectrum (black),
the model spectrum (red), and the full light-curve integrated spectrum
(blue dashed).  Note that the observed LE spectrum includes strong
nebular emission lines that are not associated with the SN.  Both the
model spectrum and the full light-curve weighted integrated spectrum
show a good fit for features blueward of 5500~\AA.  However, the lower
panel of Figure~\ref{fig:87a_specfit} shows that the spectrum redward
of 5500~\AA, and in particular the H$\alpha$ feature, are best fit by
the spectrum corresponding to the best-fit LE profile model.  From
Figure~\ref{fig:87a_example}, it is clear that the strength of the
H$\alpha$ feature in the integrated spectrum is highly dependent on
the window function.  At late times, the spectra of SN~1987A are
dominated by strong H$\alpha$ emission.  The excellent fit to the LE
spectrum of SN~1987A validates our method of determining the effective
light curve from the LE profile.

\subsection{Impact of Dust Parameters on Observed Spectrum}

\begin{figure}
\epsscale{1.2}
\plotone{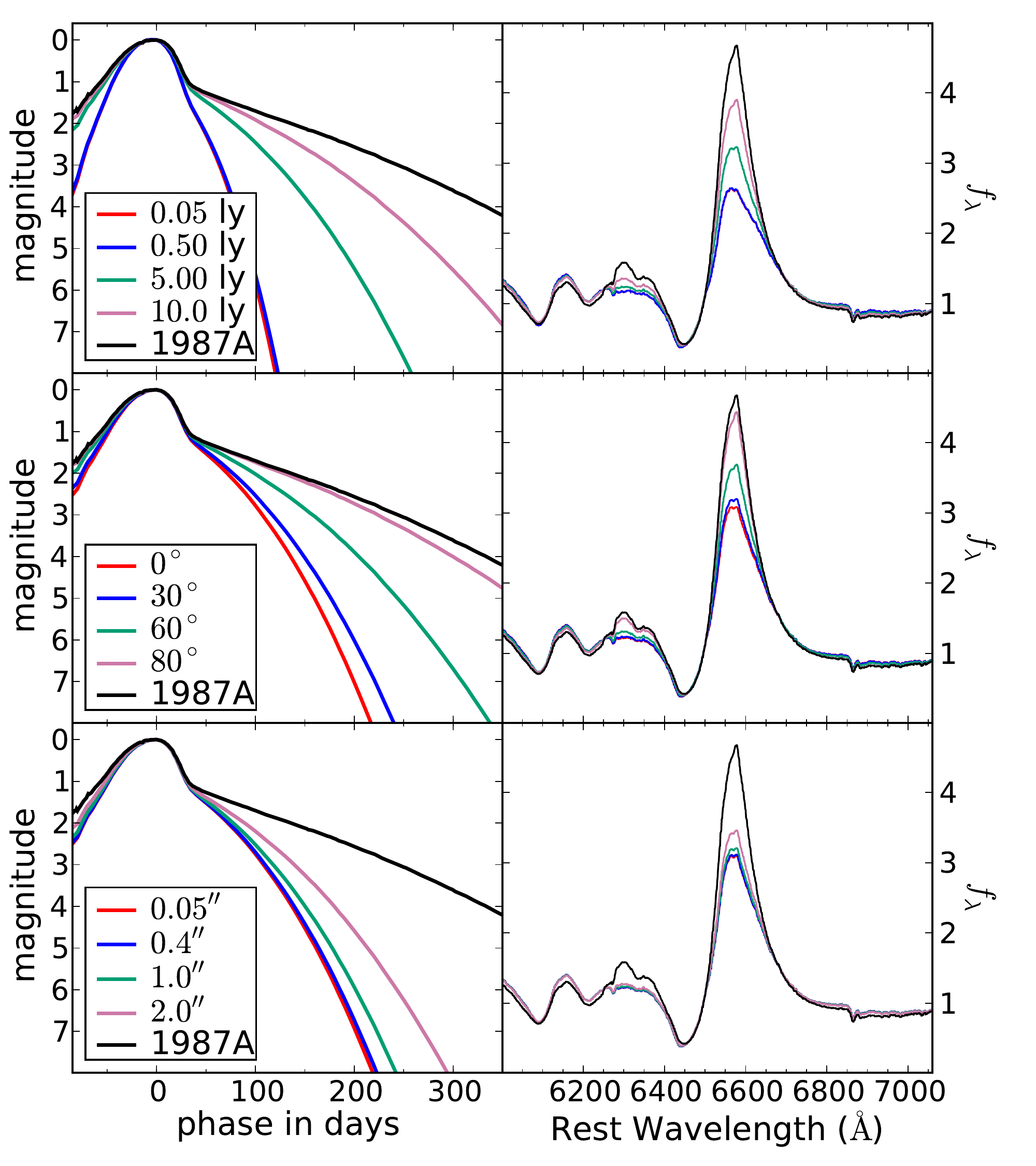}
\caption[]{Effective light curves (left column) and integrated spectra
(right column) for SN~1987A, varying the dust width, $\sigma_{d}$ (top
row), the dust inclination, $\alpha$ (middle row), and the seeing
(bottom row).  For a given row, only one parameter varies, with the
other parameters are set by the fit described above and shown in
Figure~~\ref{fig:87a_example}.
\label{fig:87a_varparams}}
\end{figure}

Similar to the analysis performed in Section~\ref{sec:systematics}, we
vary the three free parameters of our model to determine their impact
on the observed LE spectrum for SN~1987A.
Figure~\ref{fig:87a_varparams} shows the effective light curves (left
column) and integrated SN~1987A spectra (right column).  The top,
middle, and bottom rows show the model varying the dust width
($\sigma_{d}$), dust inclination ($\alpha$), and seeing, respectively,
while holding the other parameters constant.  To emphasize the
differences in the integrated spectra, the figure focuses on the
H$\alpha$ feature.  The advantage of the SN~1987A observations over
those of Cas~A is that there is a complete linking between the LE and
the final integrated spectrum via observation, since we have a priori
knowledge of the SN creating the observed LE profile.

The dependence on the integrated spectra of SN~1987A are similar to
that of SN~1993J (see Figure~\ref{fig:casa_changeparams}), where
increasing the dust width, inclination, or seeing all share the result
of stretching the weight function closer to the $wf = 1$ case,
resulting in an effective light curve approaching that of the original
SN~1987A light curve.  Figure~\ref{fig:87a_varparams} allows us to
rank the impact of the dust parameters on the observed LE spectrum.
Seeing has the least significance on the spectrum, although we note
that poor image quality should still be taken into account.  Dust
inclination has a large effect, although the sensitivity is small if
the dust is close to perpendicular to the observer.  The dust width
has the greatest impact on the spectrum, which is particularly
noteworthy since this quantity has been ignored until now. For
SN~1987A, the H$\alpha$ peak height increases by about 20\% if the
dust thickness is 10~ly instead of 5~ly, indicating a sensitivity that
exceeds that of the other parameters.

Figure~\ref{fig:87a_varparams}, as well as the SN~1987A spectrum
presented in Section~\ref{sec:87a_model}, clearly show the danger in
interpreting spectroscopic data from LEs without proper modeling.
Any interpretation of LE spectra based on relative line strengths must
be accompanied by a careful analysis of the dust properties associated
with the LE.


\section{Discussion}
\label{sec:discussion}

In the previous sections, we have presented the LE model, the
dependency of the model on various parameters, and have applied that
model to multiple real-world cases.  In this section, we discuss how
we can use our new understanding of LEs and the scattering dust to
perform additional observations that will provide new insight into
historical SNe.

\subsection{Verification of Model Parameters And Systematic Biases}

The analysis of SN~1987A LEs has the added benefit that the validity
of our dust models and their impact on the observed LE profiles and
spectra can be tested, since the spectrophotometric evolution of the
SN itself has been monitored extensively.  We show in
Section~\ref{sec:87A} and in more detail in \cite{Sinnott_87Aspec}
that the observed LE profile, apparent motion, and spectra can all be
brought in excellent accordance with what is predicted using the
SN~1987A spectrophotometric library and our dust model locally
parameterized as a planar filament with a Gaussian density profile.
This shows that at least in the case of SN~1987A, the Gaussian dust
model is a good approximation of the true dust profile.

In general, dust structures are filamentary in nature, which is easily
visible in its effect on the shapes of LEs, which often show twists
and furcations.  However, on the few arcsecond scale, most of the
shape and profiles of LEs change slowly over months and even years.
For example, the dust width, $\sigma_{d}$, for LE3923 and LE2521 is
nearly constant over 20\arcsec\ (see second panel from the bottom in
Figure~\ref{fig:casafitparams}) and smoothly increasing, respectively.
In order to account for small changes in the dust parameters, we split
the slit into 1\arcsec\ sections and fit the dust parameters
separately for each of these sections.

If there are significant deviations from the Gaussian dust density
model (e.g., if there is substructure in the density on small scales),
the window function can potentially be biased, which in turn, can
cause spectral differences that would be interpreted as intrinsic to
the SN.  We can test and guard against these systematic biases: LEs
often come in ``groups''; i.e., there are often several distinct LEs
separated by tens of arcseconds for LEs of Galactic SNe.  The
scattering dust filaments likely belong to the same dust structure,
but are tens of light years apart.  Therefore any substructure in
these filaments is uncorrelated.  For LEs of the same source event
only tens or hundred of arcseconds apart, the physical properties of
the source event are the same.  An upper limit on the systematic bias
can be set by the differences in the observed spectra which are not
accounted for by the differences in fitted dust parameters.  In all
data obtained by our team, we have not found any evidence that there
are significant biases in our analysis due to substructure in the dust
or deviation from the dust model.  This is particularly important for
the two novel methods that we describe in the next two sections.

If LEs from two different SNe are reflected off the same dust, then
the dust thickness measured from our method should be the same for
both sets of LEs.  Such a scenario is a direct test of the model fits.
Fortunately, this situation occurs for LEs from Cas~A and Tycho, and a
comparison of the derived values will be presented in a future work.
Since this situation was discovered in a relatively small and low
filling factor survey, it is reasonable to expect other such regions
of the sky exist where dust is at the intersection of two historical
SN LE ellipsoids.

\subsection{Temporally Resolving Light Echo Spectra}
\label{sec:casa_timeresolve}

One of the most exciting opportunities LEs present is to use them to
{\it temporally resolve} the SN spectrum: If a slit is aligned
perpendicular to a LE profile instead of parallel, different parts of
the slit probe different epochs of the SN light curve. 

\begin{figure*}
\epsscale{1.15}
\plottwo{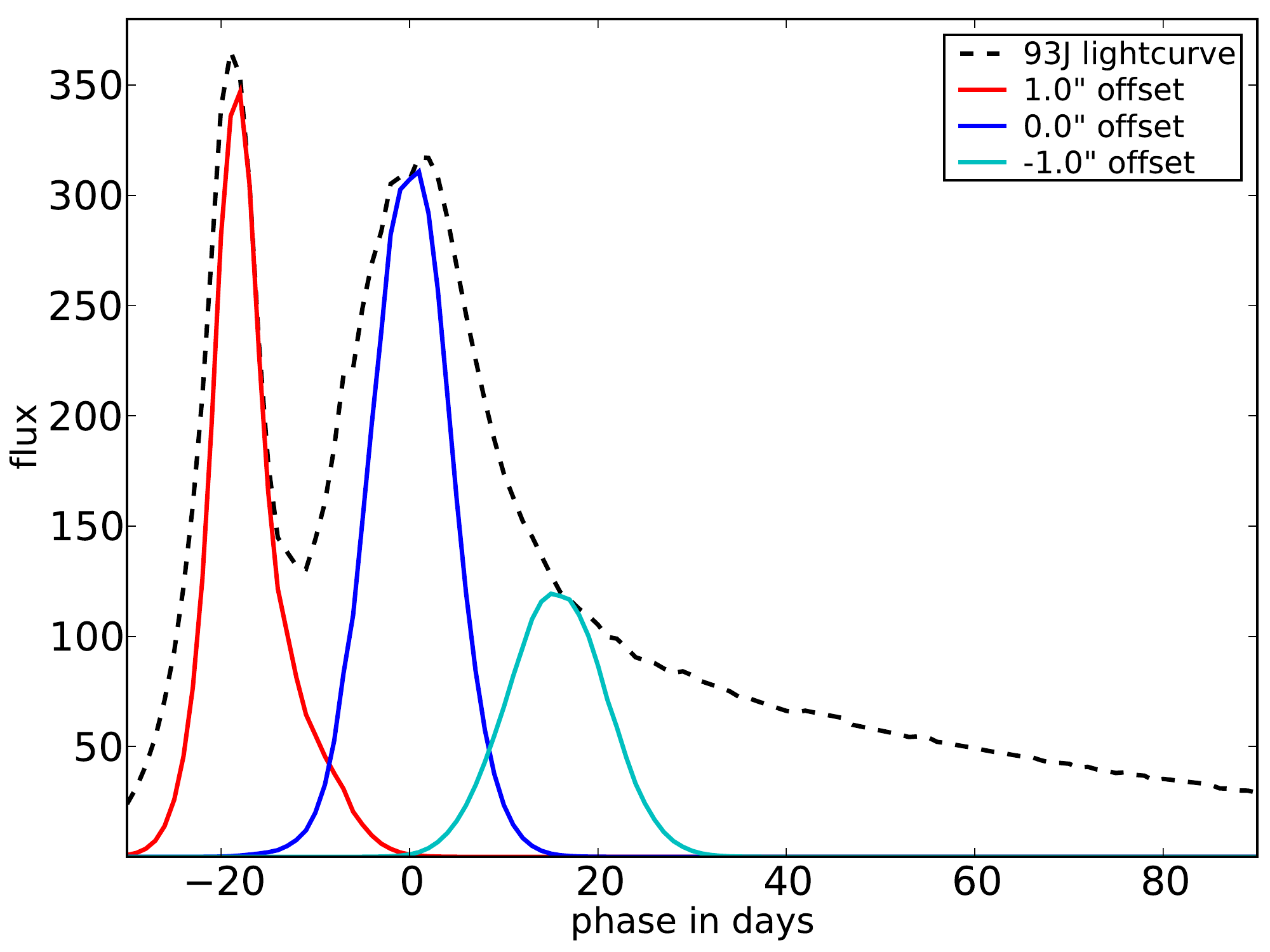}{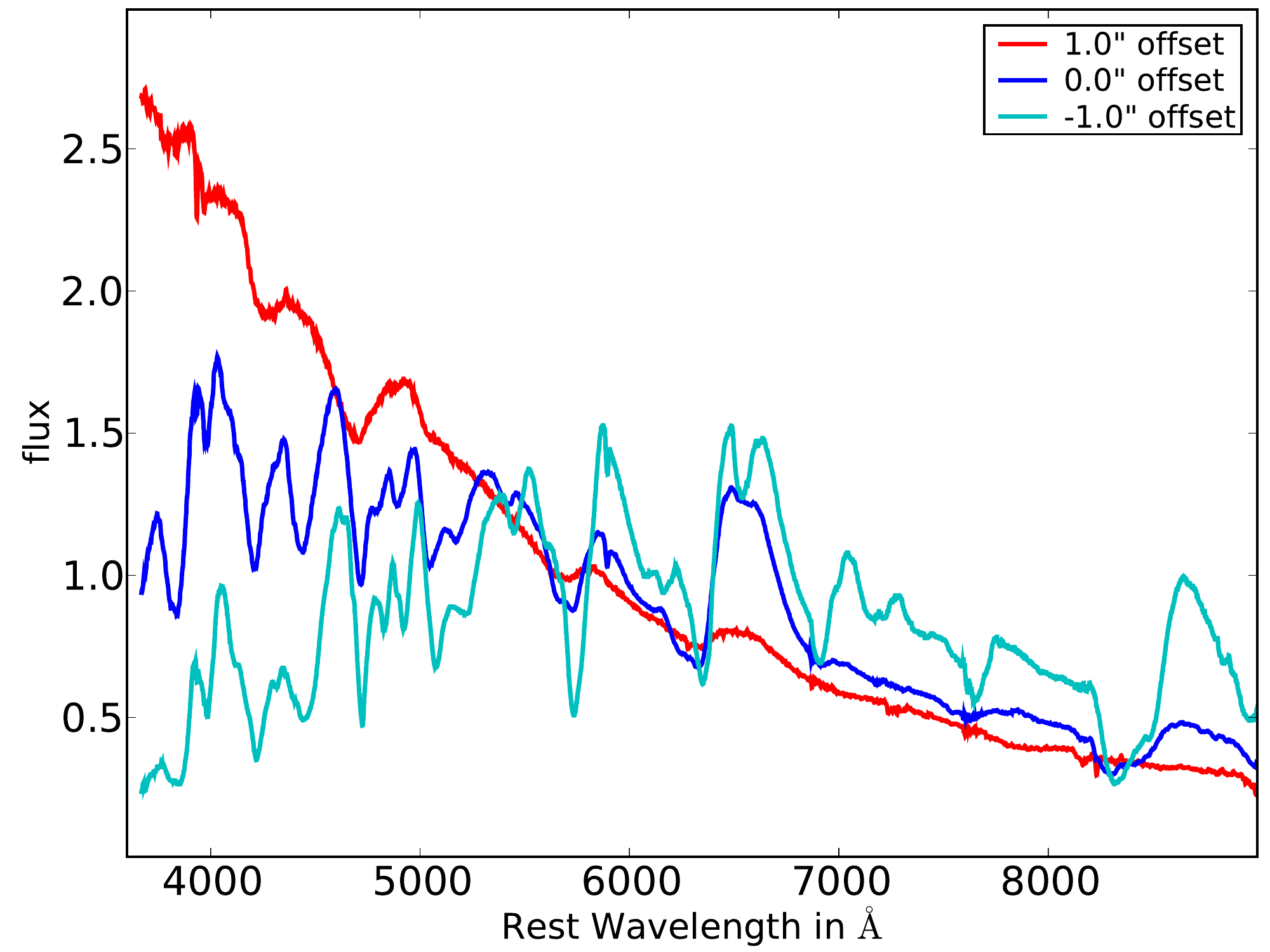}
\caption[]{\emph{Left:} Effective light curves for LE2116 with a $1\arcsec$
slit and offsets, $\Delta\rho_{\rm offset}$, of $-1.0\arcsec$,
$0.0\arcsec$, and $1.0\arcsec$ shown in red, blue, and cyan,
respectively.  \emph{Right:} Integrated spectra corresponding to the
effective light curves in the left panel.
\label{fig:casa_timeresolve}}
\end{figure*}

As an example, we imagine a SN LE where the SN is a carbon copy of
SN~1993J, the dust filament is at $z = 400$~ly with an inclination of
0\arcdeg\ and a width of 0.02~ly.  With a {\it Hubble Space
Telescope}-like ({\it HST}) PSF FWHM of $0.05\arcsec$, the projected
light curve would have minimal distortion.  If a spectrum is taken
with the slit perpendicular to the LE arc (i.e., parallel to the line
from the LE to the SNR) with $0.05\arcsec$ seeing, then the spatially
resolved spectrum (i.e., moving along the axis perpendicular to the
wavelength direction) corresponds to a temporally resolved spectrum.
The left panel of Figure~\ref{fig:casa_timeresolve} shows the
effective light curves for this situation if spectra are extracted
from $1\arcsec$ long parts of the slit.  The offset from the peak of
the LE are $1.0\arcsec$, $0.0\arcsec$, and $-1.0\arcsec$ along the
$\rho$-axis.  The right panel shows the corresponding integrated
spectra.  The three spectra show clear differences, in particular the
spectra with $\Delta
\rho_{\rm offset} = 1.0$ is very blue since it is dominated by the
first peak of the light curve.

Aligning the slit perpendicular to the LE can significantly
reduce the amount of light within the slit, and therefore, this method
is best applied to the brightest LEs.  To maximize the temporal
resolution of the LE spectra, the LE should have the following properties:
\begin{enumerate}
  \item The dust filament is very thin--- on the order of 0.01~ly.
  \item The inclination is close to the LE ellipsoid tangential.  This
  ``stretches'' the projected light curve along the $\rho$-axis.
  \item The LE is bright.
  \item The PSF of the spectroscopic observation is small.
\end{enumerate}

Even though there are no known Galactic LEs that fulfill all of these
criteria, we note that the existing surveys are far from complete.
Additionally, even in non-ideal conditions, some temporal information
could be obtained.  The ability to temporally resolve the SN spectra
will allow us to follow the evolution of the explosion, specifically
measure velocity gradients of spectral features, and potentially
probing the abundances of elements in the ejecta.

\subsection{Constraining Supernova Light-Curve Shapes}
\label{sec:casa_constraininglc}

Under similar circumstances needed to temporally resolve the LE
spectra (see Section~\ref{sec:casa_timeresolve}), we will also be able
to {\it spatially resolve} the temporal characteristics of the source
event light curve. The most important requirements are that the dust
width has to be thin, the inclination close to the LE ellipsoid
tangential, and the PSF needs to be significantly smaller than the
width of the LE profile. Note that in contrast to temporally resolving
the LE spectra, constraining the LE shape can be done with LEs that
have a significantly lower signal-to-noise ratio.

Unfortunately, there is a degeneracy between the dust width and the
light curve parameters (e.g., peak width and height; peak ratio if
double-peaked) which can be broken if different areas of the LE have
different dust widths.  Since the light curve must be the same for all
LEs in a given direction, any difference in the shape must be from the
dust.  If this degeneracy cannot be broken, LEs can still provide
interesting constraints.  First, the light curve can only be broadened
by the dust, so any profile will provide some constraint on the
light-curve shape.  Additionally, by comparing the profile to known SN
light curves, the properties of the SN can be further constrained.

The light-curve shape of a SN can give important insights into the
nature of the SN explosion.  SNe~Ia, for example, show a clear relationship
between the width of their light curves and their intrinsic brightness
\citep[e.g.,][]{Phillips93}.  Another example is if the shock
break-out of a core-collapse SN is resolved in the light curve, which
would provide a measurement (or strict constraints) on the radius of
the progenitor star.  We will discuss this in more detail in the
following paragraphs.
 
After the gravitational collapse of the core of a massive star, the
core bounces, sending a shock wave out through the outer layers of the
star, exploding the star \citep{Woosley86}.  When the shock wave
reaches the surface of the star, a flash (at optical wavelengths, it
has a thermal spectrum) is generated that decays over hours to days
\citep{Colgate74, Klein78}.  This has been seen for SNe~IIP
\citep{Gezari08, Schawinski08}, but was best observed recently with
SN~Ib~2008D \citep{Soderberg08, Modjaz09}.  As the stellar envelope
cools, the optical emission fades away.  This subsequent fading was
first detected in SN~1987A \citep[e.g.,][]{Woosley87, Hamuy88}, which
had a shock break-out cooling phase that was unusually long.  This
cooling phase has been detected for a handful of other SNe, notably
SN~1993J \citep[e.g.,][]{Richmond94}, a SN~IIb similar to Cas~A.

The duration, luminosity, and SED of the shock break-out phase depends
on a handful of parameters such as the presence of a stellar wind and
the ejecta mass \citep{Matzner99}, but is most dependent on the
progenitor radius \citep{Calzavara04}. The peak brightness of the
subsequent fading is dependent on the ejecta mass, kinetic energy, and
progenitor radius, while the timescale for the fading is proportional
to both the radius of the photosphere and its temperature
\citep{Waxman07}. 
One caveat is that interaction of the shock with shells of material
outside the star can produce prompt emission similar to a shock
breakout.

\begin{figure}
\epsscale{1.15}
\plotone{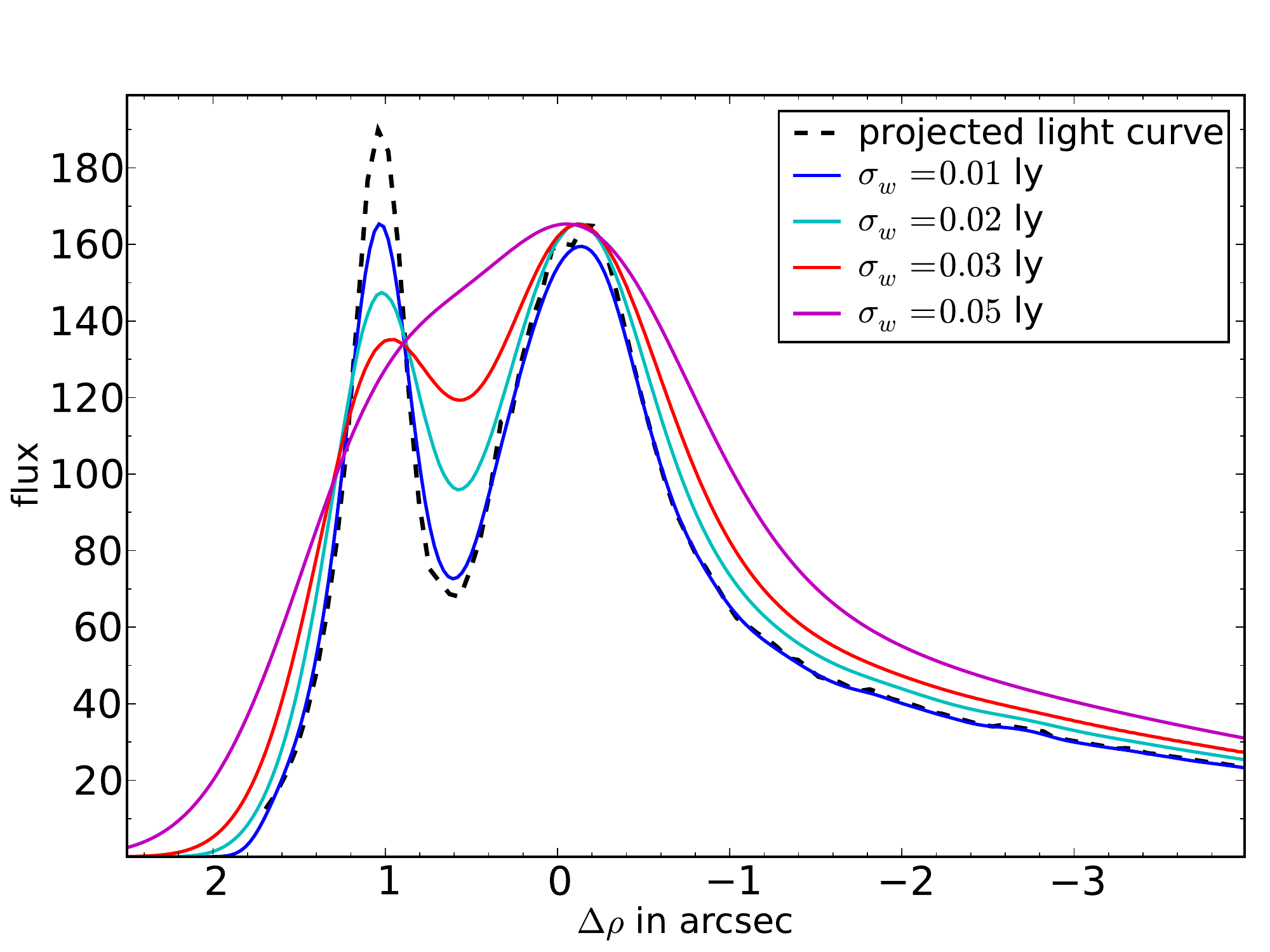}
\caption[]{Simulated LE profiles of SN~1993J assuming
different scattering dust widths.  The scattering dust filament for
the simulations is at $z = 400$~ly with an inclination of 0\arcdeg,
and the PSF FWHM is $0.05\arcsec$, similar to what can be expected
from space-based imaging.  The black dashed line is the projected
SN~1993J light curve.
\label{fig:casa_constrainlc}}
\end{figure}

The disadvantage of using LEs to observe the shock break-out is that
if the resolution of the image is poor or the dust thickness is too
large, it is impossible to distinguish between a shock break-out,
subsequent fading, and normal radioactive-decay powered SN light
curve, and a somewhat broader radioactive-decay powered SN light
curve.  For Cas~A, where we might expect to see a shock break-out, we
are unable to separate the light curve into multiple components in
ground-based images with moderate seeing ($<1$\arcsec).  However, as
shown in Figure~\ref{fig:casa_constrainlc}, if the dust thickness is
sufficiently small, we should detect a luminous shock break-out with
the resolution of {\it HST}, if it exists.  This will allow us to
constrain the radius of the progenitor and its mass loss history,
linking the physical properties of a star with its SNR.

\section{Conclusions}
\label{sec:conclusions}

In recent years, LEs of SNe have offered the rare opportunity to
spectroscopically classify the SN type hundreds of years after the
light of the explosion first reached the Earth
\citep[e.g.,][]{Rest08a}.  Recently, we have obtained the first
spectroscopic observations of a historical SN from different lines of
sight using LEs \citep{Rest10_casaspec}.  These observations provide
the unique opportunity to examine a SN from different directions, with
the light for each LE coming from slightly different hemispheres of
the SN.  In this paper, we have demonstrated that to properly separate
differences in the observed LE spectra caused by intrinsic asymmetries
in the SN from those caused by scattering dust, seeing, and
spectroscopic slit position, it is necessary to properly model the
latter.

Throughout this paper, we have shown how to both determine the
properties of the scattering dust and their influence, as well as that
of the PSF and spectroscopic slit position, on the observed LE spectra.
We found that the dust width and inclination are the dominant factors,
especially when the dust filament is thin.  These two factors can be
degenerate with respect to the LE profile shape; therefore, it is
essential to determine the dust filament inclination independently
using the LE proper motion from images taken at two (or more) epochs.
The slit width and its misalignment with the major axis of the LE are
is also a significant factor.  The image PSF smears out the observed
LE and thus also impacts the observed LE spectrum.  Taking all of
these factors into account, we found that the observed LE spectrum is
not the light-curve weighted integration of the spectra at all epochs,
but rather the integration of the spectra weighted by an effective
light curve.  This effective light curve is the original light curve
convolved with a window function, and it is different for every LE
location.  In general, the most significant difference between the
effective light curves and the original light curve was that later
epochs tend to be truncated in the effective light curves.

We used fits to the LE profiles of Cas~A and SN~1987A to test the
consistency and veracity of our model.  We fit the model to the
observed LE profiles, where the three free parameters were the LE
profile peak height, LE position, and -- physically most important --
the dust filament width.  The model reproduced the observed LE
profiles very well, and we found that it was possible to determine all
factors influencing the observed LE spectrum {\it a priori} just from
images, without using the observed spectrum.

We constructed window functions and effective light curves for the
Cas~A LEs using SN~1993J as a template. We found a much better match
of the SN~1993J spectra template to the observed Cas~A LE spectrum if
we used the effective light curve instead of the real light curve for
weighting.  The most significant difference was in the late-phase
emission lines [\ion{O}{1}] $\lambda\lambda 6300$, 6363, [\ion{Ca}{2}]
$\lambda\lambda 7291$, 7324, and the \ion{Ca}{2} NIR triplet,
consistent with our expectation from the model. Although we do not
have access to the \citet{Krause08a} images, we find that their Cas~A
spectrum is consistent with an integrated spectrum weighted with a
light curve truncated at 80~days. Our expectation is that the majority
of differences between the Cas~A LE spectrum and the full light-curve
weighted SN~1993J spectrum are due to scattering dust properties.

We also constructed a window function and effective light curve for
a SN~1987A LE which constitutes a special test case since the light
curve and spectral data of the SN are available.  With this data, we
constructed a LE spectrum from the original SN and compared it to the
observed LE spectrum.  We found that the constructed LE spectrum
correctly predicted the line strength of the H$\alpha$ line,
while a simple integration weighted by the light curve was a poor fit.

We presented several additional investigations that can be attempted
with LEs in the near future.  Two of these studies, temporally
resolved spectra and spatially resolved light curves, require thin
dust filaments.  Until now, the assumption of thick scattering dust
prevented such studies.

\acknowledgments
We thank J. Menzies for providing the spectral database of SN~1987A
obtained from SAAO. AR was partially supported by the Goldberg
Fellowship Program.  AR thanks Jeremiah Murphy for stimulating
discussions. Supernova research at Harvard College Observatory is
supported in part by NSF grant AST-0907903. Based on observations
obtained at the Gemini Observatory (program ID GS-2006B-Q-41), which
is operated by the Association of Universities for Research in
Astronomy, Inc., under a cooperative agreement with the US National
Science Foundation on behalf of the Gemini partnership: the NSF
(United States), the Science and Technology Facilities Council (United
Kingdom), the National Research Council (Canada), CONICYT (Chile), the
Australian Research Council (Australia), Minist\'{e}rio da Ci\^{e}ncia
e Tecnologia (Brazil) and Ministerio de Ciencia, Tecnolog\'{i}a e
Innovaci\'{o}n Productiva (Argentina).  NOAO is operated by AURA under
cooperative agreement with the NSF.

\bibliographystyle{fapj}
\bibliography{ms}

\appendix

\section{Light Echo Profiles}
\label{app:leprofile}

The LE equation \citep{Couderc39},
\begin{equation}
  z = \frac{\rho^2}{2ct} - \frac{ct}{2}, \label{eq:le}
\end{equation}
relates the depth coordinate, $z$, the LE-SN distance projected along
the line-of-sight, to the LE distance $\rho$ perpendicular to the line
of sight, and the time $t$ since the initial (undelayed) arrival of SN
photons.  Then the distance $r$ from the scattering dust to the source
event is $r^{2} = \rho^{2} + z^{2}$, and $\rho$ is
\begin{equation}
  \rho = (D - z) \sin \gamma, \label{eq:rhoD}
\end{equation}
where $D$ is the distance from the observer to the source event, and
$\gamma$ is the angular separation between the source event and the
scattering dust, which yields the three-dimensional position of the
dust associated with the arclet.

\citet{Tylenda04} elegantly derives the apparent expansion of a LE
ring caused by a thin plane-parallel sheet of dust.  In a very similar
manner, we derive how the dust filament inclination influences the
observed LE profile. Because of the symmetry around the $z$-axis, we
define our coordinate system so that the dust sheet is normal in the
($\rho, z$) plane.  This allows us to do all our calculations in the
two-dimensional ($\rho, z$) space.  We define the dust sheet as
\begin{equation}
  z = z_{d} - a \rho, \label{eq:dustsheet}
\end{equation}
where $a = \tan \alpha$.  We define $\alpha$ as the angle of the dust
sheet with respect to the $\rho$ axis, where positive angles go from
the positive $\rho$ axis towards the negative $z$ axis.  The dust
sheet intersects with the $z$ axis at $z = z_{d}$.  The two
intersections of the dust sheet with the reflection paraboloid defined
in Equation~\ref{eq:le} are then
\begin{align}
  \rho_{\pm}(t, z_{d}, a) &= - a c t \pm \xi, \mbox{  where} \\
  \xi &=  \sqrt{ \left ( 1 + a^{2} \right ) \left ( c t \right )^{2} + 2 z_{0} ct}.
\end{align}

Now consider a LE caused by a source event with a peak at $t_{0}$ at
an angular separation $\gamma$ from the source.  Assuming a distance
$D$ to the source event, we can then calculate the position of the
scattering dust $(\rho_{0}, z_{0})$ using Equation~\ref{eq:le} and
\ref{eq:rhoD}.  The dust sheet defined in Equation~\ref{eq:dustsheet}
goes through ($\rho_{0}, z_{0}$) if
\begin{equation}
  z_{d} = z_{0} + a \rho_{0}.
\end{equation}

There are two special dust inclinations.  We denote the inclination
angle of the dust sheet tangential and orthogonal to the LE ellipse as
$\alpha_{\parallel}$ and $\alpha_{\perp}$, respectively.  Then the
slopes of the dust sheets are $a_{\parallel} = \tan{
\alpha_{\parallel} }$, and $a_{\perp} = \tan{ \alpha_{\perp} }$, and
we can calculate them as
\begin{align}
  a_{\parallel} &= - \left ( 1 + \frac{2 z_{0}}{ct} \right )^{1/2} = - \frac{\rho_{0}}{ct}; \\
  a_{\perp} &= -a_{\parallel}^{-1} = \left ( 1 + \frac{2 z_{0}}{ct} \right )^{-1/2} = \frac{ct}{\rho_{0}}.
\end{align}
The intersection of the dust sheet with the LE ellipsoid is then at
\begin{equation}
  \rho_{0} = 
  \begin{cases}
    0 & \mbox{for } \alpha = 90\arcdeg \\
    \rho_{+} & \mbox{for } a > a_{\parallel} \\
    \rho_{-} & \mbox{for } a < a_{\parallel}
  \end{cases}. \label{eq:v}
\end{equation}

Now consider a dust sheet with a finite width defined by a density
$D(\Delta r)$, and a light curve $f_{\rm source} (\Delta t)$, where
$\Delta r$ is the distance orthogonal to the dust sheet, and $\Delta t
= t - t_{0}$ (see Section~\ref{app:dustmodel} for a more detailed
description of the dust model $D(\Delta r)$). The flux from the source
event at time $t$ is reflected by the part of the dust sheet at a
distance $\Delta r$ to the center of the dust sheet, and makes a
relative contribution to the LE signal of
\begin{equation}
  f_{\rm LE} = D(\Delta r) f_{\rm source} (\Delta t) \label{eq:fle}
\end{equation}
at the position
\begin{equation}
  \rho_{\rm LE} =
  \begin{cases}
    \rho_{+} (t_{0} + \Delta t,z_{d} + \Delta r/\cos{\alpha}, a) & \mbox{if } \rho_{0} = \rho_{+} \\
    \rho_{-} (t_{0} + \Delta t,z_{d} + \Delta r/\cos{\alpha}, a) & \mbox{if } \rho_{0} = \rho_{-}
  \end{cases}. \label{eq:rho_le}
\end{equation}

The reason for choosing $+$ and $-$ is based on the choice made for
$\rho_{0}$.  For all practical purposes, $\Delta r$ and $\Delta t$ are
small, and therefore the sign should be the same.  Exceptions are
$\alpha \approx 0.0$ and $a \approx a_{\parallel}$. We define
\begin{equation}
  \Delta \rho = \rho_{\rm LE} - \rho_{0}. \label{eq:deltarho}
\end{equation}
By examining the relative contribution for all values of $\Delta r$
and $\Delta t$ and using Equation~\ref{eq:fle} -- \ref{eq:deltarho},
we can calculate a cumulative grid of fluxes, $F^{\prime}_{\rm LE}
(\Delta t, \Delta \rho)$.

The model must now account for seeing.  We describe the data PSF as a
Gaussian with a FWHM of $x$ arcsec and Gaussian width of $\sigma_{\rm
PSF} = x/(2 \sqrt{2 \ln (2)}) \approx x/2.35$.  Then the cumulative
grid of fluxes convolved with the seeing is
\begin{equation}
  F_{\rm LE}(\Delta t, \Delta \rho) = \sum_{\Delta \rho^{\prime}} F^{\prime}_{\rm LE} (\Delta t, \Delta \rho^{\prime}) \frac{1}{\sqrt{ 2 \pi \sigma_{\rm PSF}^{2}}} \exp{ \left ( \frac{(\Delta \rho - \Delta \rho^{\prime})^{2}}{2 \sigma_{\rm PSF}^{2}} \right ) }.
\end{equation}
Integrating over $\Delta t$ returns the LE flux profile with respect
to $\rho$:
\begin{equation}
  P_{\rm LE}(\Delta \rho) = \int F_{\rm LE} (\Delta t, \Delta \rho) \mathrm{d}t. \label{eq:P_LE}
\end{equation}

\section{Dependence of the Light Echo Spectrum on Light Echo Profile}
\label{app:lespec}

We now examine the LE flux for a spectroscopic observation.  Consider
a slit positioned at $\rho_{s}$ with a effective width of $\sigma_{s,
{\rm eff}}$.  The reason that this is the effective slit width and its
relation to the real, physical slit width $\sigma_{s}$ is explained
below in Appendix~\ref{app:slitmisalignement}.  We define the offset
between the slit and the LE measured parallel to the $\rho$ axis as
$\Delta \rho_{\rm offset}= \rho_s - \rho_{0}$.  The slit then subtends
from
\begin{align}
  \Delta \rho_{\min} &= \Delta \rho_{\rm offset} - \sigma_{s, {\rm eff}}/2 \mbox{  to} \label{eq:rhomin} \\
  \Delta \rho_{\max} &= \Delta \rho_{\rm offset} + \sigma_{s, {\rm eff}}/2. \label{eq:rhomax}
\end{align}
We can then calculate the relative contribution, $H(\Delta t)$, of the
source event at time $\Delta t$ to the spectra with
\begin{equation}
  H(\Delta t) = \int_{\Delta \rho_{\min}}^{\Delta \rho_{\max}} F_{\rm LE}(\Delta t, \Delta \rho) \mathrm{d}\Delta \rho. \label{eq:H}
\end{equation}
$H(\Delta t)$ is also the effective light curve.  We define the window
function $w(\Delta t)$ as the fractional difference between the
effective light curve and the real light curve,
\begin{equation}
  w(\Delta t) = f_{\rm source}(\Delta t) / H(\Delta t). \label{eq:Hw}
\end{equation}
The observed LE spectrum $S_{\rm LE}(\lambda)$ is then 
\begin{equation}
  S_{\rm LE}(\lambda) = \int S(\lambda, \Delta t) H(\Delta t) \mathrm{d}\Delta t, \label{eq:lespec}
\end{equation}
where $S(\lambda, \Delta t)$ is the spectrum of the source event at
the given phase $\Delta t$.

\section{Dependence of the Light Echo Spectrum on Slit Position}
\label{app:slitmisalignement}

The position and inclination of the slit with respect to the LE has a
major impact on the observed spectrum.  We illustrate this in
Figure~\ref{fig:slitpos_illu}.  We define a coordinate system ($\rho,
u$), where $u$ is orthogonal to $\rho$ through $\Delta \rho = 0.0$ in
the plane of the sky.  For a dust filament that is locally planar and
perpendicular to the ($\rho, z$) plane (i.e., parallel to $u$), the LE
is aligned with the $u$ axis.  However, if it is tilted with respect
to $u$ by an angle of $\delta_{1}$, then the LE will also be tilted by
the same angle (see gray shaded area in
Figure~\ref{fig:slitpos_illu}).

Now consider a slit (red box in Figure~\ref{fig:slitpos_illu}) that is
misaligned with the LE by an angle $\delta_{2}$, and also offset along
the $\rho$ axis by $\Delta s$. We split the slit into subslits,
indicated with the dotted blue lines in Figure~\ref{fig:slitpos_illu}.
For each of these subslits, we determine the LE profile in
Figure~\ref{fig:slitpos_illu}.  The offset $\Delta \rho_{{\rm offset},
i}$ of the subslit $i$ centered at $u_{i}$ to the LE is
\begin{equation}
  \Delta\rho_{{\rm offset}, i} = u_{i} \sin(\delta_{2}) + \Delta s, \label{eq:delta_rho_offset}
\end{equation}
where $\delta_{2}$ is the angle between the slit and the LE, and
$\Delta s$ is the offset of the center of the slit to the LE (see
Figure~\ref{fig:slitpos_illu}).  The effective slit width $\sigma_{s,
{\rm eff}}$ depends on the physical slit width $\sigma_{s}$ and is
\begin{equation}
  \sigma_{s,{\rm eff}} = \sigma_{s}/\cos(\delta_{2}). \label{eq:slitwidth_eff}
\end{equation}

The effective light curve $H_{i}$ and window function $w_{i}$ for each
subslit $i$ can be calculated plugging $\Delta \rho_{{\rm offset}, i}$
into Equation~\ref{eq:rhomin} and \ref{eq:rhomax}.  The total window
function $w_{\rm slit}$ for the full slit is then the weighted average
of the subslit window functions, where the weight is the flux $f_{w}$
that falls into the subslit is
\begin{align}
  f_{w, i} &= \int_{\Delta \rho_{\min}}^{\Delta \rho_{\max}} P_{\rm LE}(\Delta \rho) \mathrm{d}\Delta \rho \mbox{  with}\\
  w_{\rm slit} &= \frac{\sum_{i} w_{i} f_{w, i}^{-2}} {\sum_{i} f_{w, i}^{-2}}. \label{eq:wslit}
\end{align}

The integrated spectrum can then be calculated by using $w_{\rm slit}$
in Equations~\ref{eq:Hw} and \ref{eq:lespec}.

\section{Dust Model}
\label{app:dustmodel}

For simplicity in our simulations, we choose a box-car dust density
profile $D(\Delta r)$.  A more realistic density profile is a
Gaussian, and therefore when fitting observed LE profiles we use a
profile
\begin{equation}
D(\Delta r) = \frac{c}{\sqrt{2 \pi \sigma_{d}^{2}}} \exp{ \left ( \frac{\Delta r^{2}}{2\sigma_{d}^{2}} \right )},
\end{equation}
where $\sigma_{d}$ is the width of the Gaussian, and $c$ is a
normalization factor.  Note that the physical dust density profile
width $\sigma_{d, {\rm phys}}$ is different from the nominal fitted
$\sigma_{d}$ if $\delta_{1} \ne 0$.  They are related through
\begin{equation}
  \sigma_{d, {\rm phys}} = \sigma_{d} \cos(\delta_{1}).
\end{equation}
%

\section{Dust Filament Inclination}
\label{app:inclination}

We refer the reader to a more thorough and detailed discussion about
the apparent motion and its dependence on the dust filament
inclination by Rest et~al.\ (in prep.).  However, we present an
initial treatment of the process here.  We define the dust filament 
inclination $\alpha$ as
\begin{equation}
  \alpha = \tan^{-1} \left (-\frac{z(t_{2}) - z(t_{1})}{\rho(t_{2}) - \rho(t_{1})} \right )
\end{equation}
with $t_{2} > t_{1}$, and with $z$ calculated from
Equation~\ref{eq:le} for a measured $\rho$ and given $t$.  Note that a
dust filament in the plane of the sky has $\alpha = 0\arcdeg$, and
dust filaments tilted away from the observer have positive $\alpha$.
If more than two epochs are available, fitting a straight line through
($\rho,z$) increases the accuracy of $\alpha$.

There are several sources that can introduce errors into the
measurement of the dust inclination, specifically:
\begin{itemize}
  \item Poisson uncertainty in the apparent motion of the LE,
  especially for the fainter LEs
  \item A small number of epochs
  \item Intrinsic differences in the filament inclination with $\rho$
\end{itemize}
The first two mainly increase the Poisson noise in the
inclination. The last one, however, is the most significant one since
it can introduce systematic uncertainties. If the dust filament has
additional substructure, such as twisted dust on small scales, then
the inclination of the scattering dust changes on small scales and a
correct constraint on the dust inclination at the time of spectroscopy
can only be done if there are several epochs close to the time of the
spectroscopy.


\end{document}